\documentclass[fleqn,usenatbib]{mnras}

\usepackage{newtxtext,newtxmath}

\usepackage[T1]{fontenc}

\DeclareRobustCommand{\VAN}[3]{#2}
\let\VANthebibliography\thebibliography
\def\thebibliography{\DeclareRobustCommand{\VAN}[3]{##3}\VANthebibliography}

\usepackage{graphicx}	
\usepackage{amsmath}	
\usepackage{xcolor}     

\newcommand{\gro}{GRO~J1655-40}	

\title[
What powers the disc wind in GRO J1655-40?]
{
What powers the wind from the black hole accretion disc in GRO J1655-40?
}

\author[Tomaru \& Done, Mao]{
Ryota Tomaru$^{1}$\thanks{E-mail: ryota.tomaru@durham.ac.uk}, 
Chris Done$^{1}$\thanks{E-mail: chris.done@durham.ac.uk}
Junjie Mao$^{2,3}$\thanks{E-mail: mao@astro.hiroshima-u.ac.jp}, 
\\
$^{1}$Centre for Extragalactic Astronomy, Department of Physics, University of Durham, South Road, Durham DH1 3LE, UK\\
$^{2}$Department of Physics, Hiroshima University, 1-3-1 Kagamiyama, Higashi-Hiroshima, 739-8526, Hiroshima, Japan\\
$^{3}$SRON Netherlands Institute for Space Research, Niels Bohrweg, Leiden, 2333 CA, the Netherlands.
}

\date{Accepted XXX. Received YYY; in original form ZZZ}

\pubyear{2022}

\begin{document}

\label{firstpage}
\pagerange{\pageref{firstpage}--\pageref{lastpage}}
\maketitle

\begin{abstract}
Black hole accretion discs can produce powerful outflowing plasma (disc winds), seen as blue-shifted absorption lines in stellar and supermassive systems.
These winds in Quasars have an essential role in controlling galaxy formation across cosmic time, but there is no consensus on how these are physically launched.
A single unique observation of a stellar-mass black hole GRO J1655-40 
was used to argue that magnetic driving was the only viable mechanism and motivated unified models of magnetic winds in both binaries and Quasars.  The alternative, X-ray heating (thermal-radiative wind), was ruled out for the low observed luminosity by the high wind density estimated from an absorption line of a metastable level of Fe {\sc xxii}.
Here we reanalyse these data using a photoionisation code that includes cascades from radiative excitation
as well as collisions in populating the metastable level.
The cascade reduces the inferred wind density by more than an order of magnitude. 
The derived column is also optically thick, so the source is intrinsically more luminous than observed.
We show that a thermal-radiative wind model calculated from a radiation hydrodynamic simulation matches well with the data.
We revisit the previous magnetic wind solution and show that this is also optically thick,
leading to a larger source luminosity. However, unlike the thermal-radiative wind, it struggles to reproduce the overall ion population at the required density.
These results remove the requirement for a magnetic wind in these data and remove the basis of the self-similar unified magnetic wind models extrapolated to Quasar outflows.
\end{abstract}




\section{Introduction}
Black hole accretion flows are generically able to launch winds. These are seen in Active Galactic Nuclei (AGN), where the mass,
momentum and energy carried by the wind impact the host galaxy, controlling star formation via AGN feedback \citep{Crenshaw2003,Laha2021}.
If this wind is magnetic then it can also contribute to (or even dominate) the angular momentum transport process in the disc.
A small amount of material accelerated along the field lines can carry most of the angular momentum outwards, enabling accretion by allowing the rest of the material to fall inwards \citep{Blandford1982}.
Such magnetic winds are scale free, so should be present in the much brighter Galactic binary systems as well as in AGN. 

Winds are seen as blue-shifted absorption features in the Galactic binaries \citep{Ueda2004,Miller2006, Miller2006b,Miller2016}. 
However, unlike AGN, where there are multiple other wind launching mechanisms due to their lower disc temperatures, 
the only competing process in binaries is thermal winds, driven by X-ray heating of gas in the photosphere of the outer disc, or thermal-radiative winds, where radiation pressure gives an additional component to the outwards acceleration.
The launch radius of thermal winds is determined by the spectrum of the illuminating radiation. This heats the photosphere to an equilibrium temperature, $T_{\rm IC,7}$, produced by Compton heating/cooling and given in units of $10^7$~K. This is independent of radius, while the escape velocity due to gravity decreases as $v_{\rm esc} = 5\times 10^{2}~T_{\rm IC,7}^{0.5} ~{\rm km~s^{-1}}$. This means the irradiated gas is unbound for radii above
$R_{\rm IC}=6\times10^{5}/T_{\rm IC,7}\ R_g$. This dependence of the wind properties (launch radius/velocity) on the Compton temperature mean that thermal winds respond
structurally to changes in the illuminating spectrum, in addition to the predicted change in ionisation.

Observationally, winds are
only seen in a very specific subset of objects. These all at fairly high inclination angles and 
have soft (disc dominated) spectra, with $T_{\rm IC,7}\sim 1$ (\citealt{ponti2012}). They have long orbital periods i.e. large discs
as required for the thermal winds at these irradiating temperatures, and the wind velocity is of order 
$100-1000~{\rm km~s^{-1}}$\citep{DiazTrigo2016}, again as predicted by the thermal wind models.
This suggests that thermal driving is the main mechanism which produces the observed winds in X-ray binaries. 

Nonetheless, radiation driving should also modify the resulting thermal wind structure for $L/L_{\rm Edd}>0.1-0.2$. This is confirmed by 
recent radiation hydrodynamics simulations
\citep{Tomaru2019, Higginbottom2020}. 
These show that including radiation driving in the thermal wind calculations (thermal-radiative winds) is important to get a detailed match to the observations, otherwise the winds seen at high inclination are
slower than observed (see also e.g. \citealt{Luketic2010,Higginbottom2014}). The radiation force is not just from electron scattering as the 
wind is not completely ionised close to the disc surface in the disc dominated soft states, so there is some remaining bound-free (edges) and bound-bound (line) opacity. All three components add together to the total radiation force which helps unbind the material when thermal heating alone is insufficient. Results from these simulations can even 
match in detail to the observed 
absorption lines from He and H-like Fe in typical objects (e.g. H1743-322 in \citealt{Tomaru2020} and GX 13+1 in \citealt{Tomaru2020b}). 

The simulations also confirm that
thermal winds should become invisible when the source makes a transition to the hard state, with $T_{\rm IC,7}\sim 10$. The wind may be launched
quite efficiently (though this depends on details of the inner disc geometry) but it is faster, so is less dense. This, together with the harder spectral illumination, means that it is completely ionised so there is no radiation force from bound-bound and bound-free, and these states are typically seen at $L/L_{\rm Edd}<0.1$ so electron scattering is not important either
(\citealt{Tomaru2019}).

Despite these successes of the
thermal-radiative wind models, an alternative mechanism, magnetic driving,
is still considered because of 
a single observation of a very 
unusual wind in GRO J1655-40. 
The high-quality X-ray grating spectrum by {\it Chandra} of GRO\,J1655-40 during its 2005 outburst shows absorption lines from unusually low ionisation states, including a density sensitive metastable absorption line ($\lambda 11.92$ of B-like Fe {\sc xxii}). This gives a wind density of $(4.0-6.3)\times10^{13}~{\rm cm^{-3}}$ (\citealt{Miller2008}, revised down from the original claim of $5.6\times10^{15}~{\rm cm^{-3}}$ in \citealt{Miller2006}). This gives a direct estimate of the launch radius when combined with the observed luminosity, $L\sim 0.04L_{\rm Edd}$, and ionisation parameter
$\xi=L/(nR^2) \sim 10^{5}~{\rm erg ~cm ~s^{-1}}$ \citep{Miller2006}
, where $n$ is the number of densities for hydrogen and $R$ is the radius from the source
(In this paper, we use both $R$ and $r$ to show radii. 
For the model of thermal winds in Sec. 4 we use $R$. 
For the  model of magnetic winds in Sec.5 we use $r$. 
In Sec.5, $r_d$ shows the radius along the disc mid-plane and $r_0$ shows the inner radius of this object taken from \citealt{Fukumura2017} ). 
The resulting radius is more than a factor 10 smaller than expected for a thermal wind, 
{\em requiring} magnetic launching \citep{Miller2006,Miller2008}. 
By extension, this supported a unified model of magnetic winds
across the mass scale, including both stellar and supermassive black holes (\citealt{Fukumura2017}, hereafter F17).

However, the density derived above assumed that the metastable level is populated only by collisional (electron impact) excitation but photoexcitation can also be important \citep{Mauche2003, Mauche2004}. Electrons in the ground level can be photo-excited to upper levels, and 
de-excite via a radiative cascade which can populate the 
metastable level. This process is not included in the standard photo-ionisation code used to analyse X-ray data ({\sc xstar}: \citealt{Kallman2001}). 
Here instead we use the the photoionisation code {\tt pion} \citep{Miller2015, Mehdipour2016, Mao2017} in the X-ray spectral analysis code {\sc spex} \citep[v3.05,][]{Kaastra2018} to include these processes and re-assess the density of the photoionized wind in GRO\,J1655-40 in a self-consistent way. This significantly reduces the inferred density, bringing the launch radius back into the range of thermal-radiative winds. Additionally, the derived column density from this code implies that the wind is optically thick so that the intrinsic luminosity is much larger than the observed luminosity. 

We show an explicit optically thick thermal-radiative wind model which gives a good overall match for the wind properties,
aligning this critical observation with all other data from binary winds 
for which the thermal-radiative wind simulations provide a good match \citep{Tomaru2019,Tomaru2020,Tomaru2020b,Higginbottom2019,Higginbottom2020}. 

Our result removes the requirement for magnetic winds in binaries. We also show some specific 
magnetic wind models, based on the self similar solutions of F17 and show that these
also go optically thick, and cannot easily match even our new lower density, let alone the original density which motivated the magnetic wind models in the first place. This 
challenges the plausibility of magnetic models to explain these data, 
and so by extension, the argument  for self-similar magnetic winds in AGN. 
Instead, this opens the way to a quantitative model for AGN feedback  using the known - but complex - physics of UV line driving and dust driving in supermassive black hole accretion flows.

\section{Observational data}
\label{sct:obs_data}
GRO J1655-40 is a well studied Galactic low-mass X-ray binary \citep{Hjellming1995, Abramowicz2001, Orosz1997}, 
whose distance is $d \sim 3.2 ~\mathrm{kpc}$ \citep{Hjellming1995},
orbital period is 2.6 days \citep{Greene2001}, and black hole mass is
$M_\mathrm{BH} = 5.4\pm0.3~M_\odot$ while the companion star has mass $M_s=1.45\pm 0.35~M_\odot$ \citep{Beer2002}.
The orbital period and the masses gives an outer disc radius of 
$R_\mathrm{out}= 4\times 10^{11}~\mathrm{cm} = 5\times 10^{5}~R_{g}$, where $R_{g} = GM_{\rm BH}/c^2$ assuming that the outer disc extends to 80\% of the Roche lobe radius. 

During the outburst in 2005, this object showed 
an unusual continuum shape called a hyper-soft state \citep{Neilsen2016}. This is accompanied by unusual absorption line features from low ionisation ion species (e.g., Li-, Be-, and B-like Fe ions) observed with the high energy grating (HEG) aboard the {\it Chandra} X-ray Observatory.
None of the other observations of black hole binaries shows these absorption lines \citep{Miller2006, Miller2008}.

We take these data (ObsID: 5461) on 2005-04-01 (i.e., Modified Julian Date $=53461$) from the Chandra Transmission Grating Data Catalog and Archive  \citep[TGCat, ][]{huenemoerder2011}.
Similar to previous works \citep{Miller2008, Kallman2009}, the positive and negative first-order HEG (high-energy grating) data were combined using CIAO v4.13 \citep{fruscione2006}.
In addition, the spectral data and response files are converted into {\sc spex} formats using \textit{trafo}.


\section{Photoionisation modelling of the high-resolution X-ray spectrum}

\begin{table}
\centering
\caption{Best-fit parameters of Chandra HEG spectrum of GRO J1655-40 observed on 2005-04-01. Expected $C$-statistics are calculated as described in \citet{Kaastra2017}.
Frozen parameters are indicated with (f).
All quoted errors refer to statistical uncertainties at the 68.3\% confidence level. }
\label{tbl:pars_sed}
\begin{tabular}{l|llll}
\noalign{\smallskip}
\hline
\noalign{\smallskip}
Model & SED~1 & SED~2  \\
\noalign{\smallskip}
\hline
\noalign{\smallskip}
\multicolumn{3}{c}{Statistics} \\
\noalign{\smallskip}
\hline
\noalign{\smallskip}
$C_{\rm stat}$ & 12587.0 & 12651.1 \\
$C_{\rm expt}$ & $2793\pm75$ & $2793\pm75$  \\
d.o.f. & 2775 & 2776 \\ 
\noalign{\smallskip}
\hline
\noalign{\smallskip}
\multicolumn{3}{c}{Galactic absorption} \\
\noalign{\smallskip}
\hline
\noalign{\smallskip}
$N_{\rm H}$ ($10^{21}~{\rm cm^{-2}}$) & $5.55\pm0.07$ & $5.61\pm0.02$  \\
\noalign{\smallskip}
\hline
\noalign{\smallskip}
\multicolumn{3}{c}{Comptonization} \\
\noalign{\smallskip}
\hline
\noalign{\smallskip}
Norm (${\rm ph~s^{-1}~keV^{-1}}$) & $7.8_{-2.6}^{+0.2}\times10^{46}$ & $24.0_{-3.6}^{+1.1}\times10^{46}$ \\
$T_{\rm seed}$ (keV) & $0.192\pm0.004$ & 0.193 (f)  \\
$T_{\rm plasma}$ (keV) & $1.109\pm0.003$ & $1.111\pm0.002$ \\
$\tau$ & $12.68\pm0.07$ & $12.60\pm0.09$  \\
\noalign{\smallskip}
\hline
\noalign{\smallskip}
\multicolumn{3}{c}{{\tt pion} \#1}  \\
\noalign{\smallskip} 
\hline
\noalign{\smallskip}
$N_{\rm H}~({\rm cm^{-2}})$ & $2.49_{-0.73}^{+0.04}\times10^{24}$ & $4.27_{-0.14}^{+0.10}\times10^{24}$  \\
\noalign{\smallskip}
$\log \xi~{\rm (erg~s^{-1}~cm)}$ & $3.90\pm0.01$ & $4.09\pm0.01$ \\
$f_{\rm cov}$ & $0.77_{-0.10}^{+0.01}$ & $0.92\pm0.06$ \\
$v_{\rm out}~({\rm km~s^{-1}})$ & $-598\pm16$ & $-621\pm10$ \\
$v_{\rm mic}~({\rm km~s^{-1}})$ & $143\pm7$ & $145\pm4$ \\
\noalign{\smallskip}
\hline
\noalign{\smallskip}
\multicolumn{3}{c}{{\tt pion} \#2}  \\    
\noalign{\smallskip} 
\hline
\noalign{\smallskip}
$N_{\rm H}~({\rm cm^{-2}})$ & $(4.7\pm0.4)\times10^{23}$ & $(4.7\pm0.3)\times10^{23}$  \\
\noalign{\smallskip}
$\log \xi~{\rm (erg~s^{-1}~cm)}$ & $3.38\pm0.02$ & $3.44\pm0.15$  \\
$f_{\rm cov}$ & $0.57\pm0.02$ & $0.59\pm0.01$ \\
$n_{\rm H}~({\rm cm^{-3}}$) & $(3.0\pm0.5)\times10^{12}$ & $(2.8\pm0.4)\times10^{12}$   \\
$v_{\rm out}~({\rm km~s^{-1}})$ & $-310\pm12$ & $-323\pm3$ \\
$v_{\rm mic}~({\rm km~s^{-1}})$ & $93\pm3$ & $99\pm2$ \\
\noalign{\smallskip}
\hline
\end{tabular}
\end{table}
\label{sct:obs}
\begin{figure*}
    \centering
    \includegraphics[width=0.9\hsize]{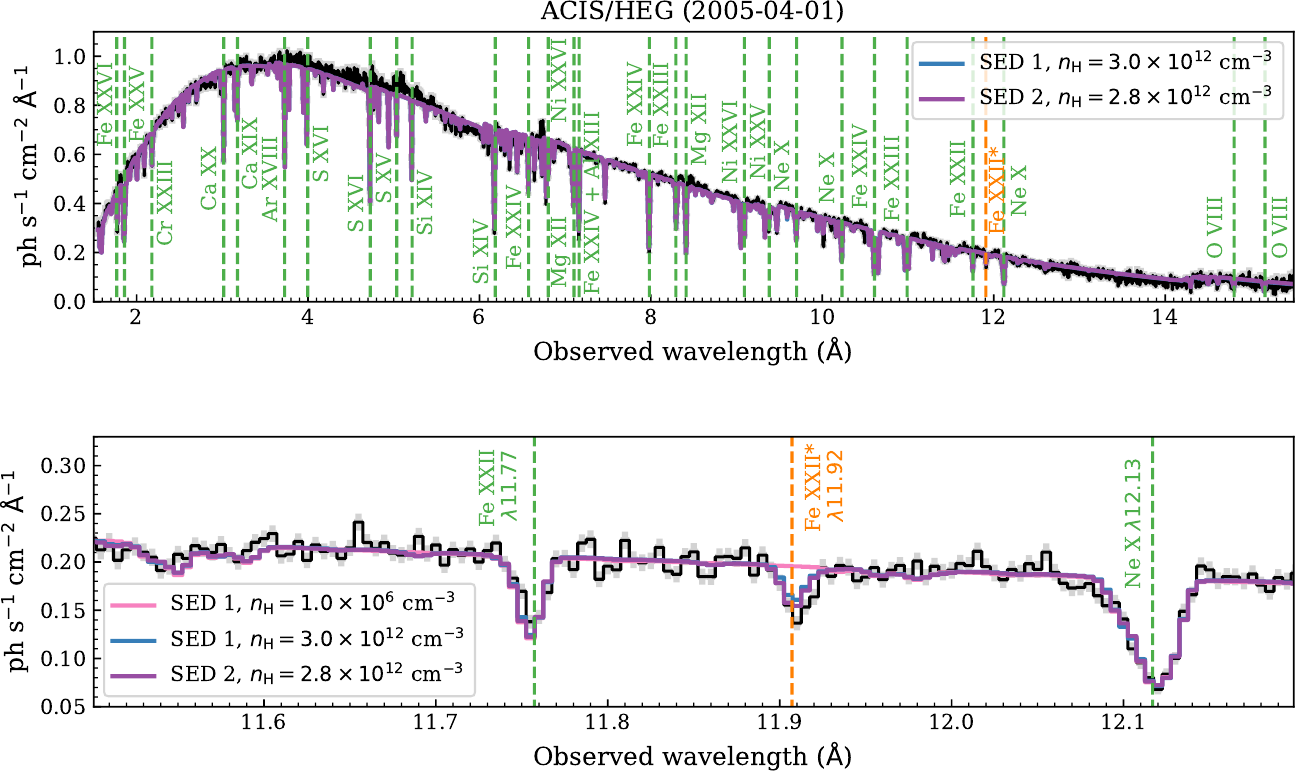}\\
    \caption{
    Photoionization modelling to the Chandra grating spectrum of \gro.
    Results from two distinct spectral energy distributions (SED) are shown (overlapped) with barely visible differences.
    The bottom panel zooms in to show the Fe {\sc xxii} and Fe {\sc xxiii} absorption lines (Table A2).
    The colours show absorption lines from ground (green) and metastable (orange) levels.
    A low-density model (pink) can not produce the observed metastable absorption lines.
    }
    \label{fig:plot_dnma_zoom}
\end{figure*}
\begin{figure*}
    \centering
    \includegraphics[width=0.9\hsize]{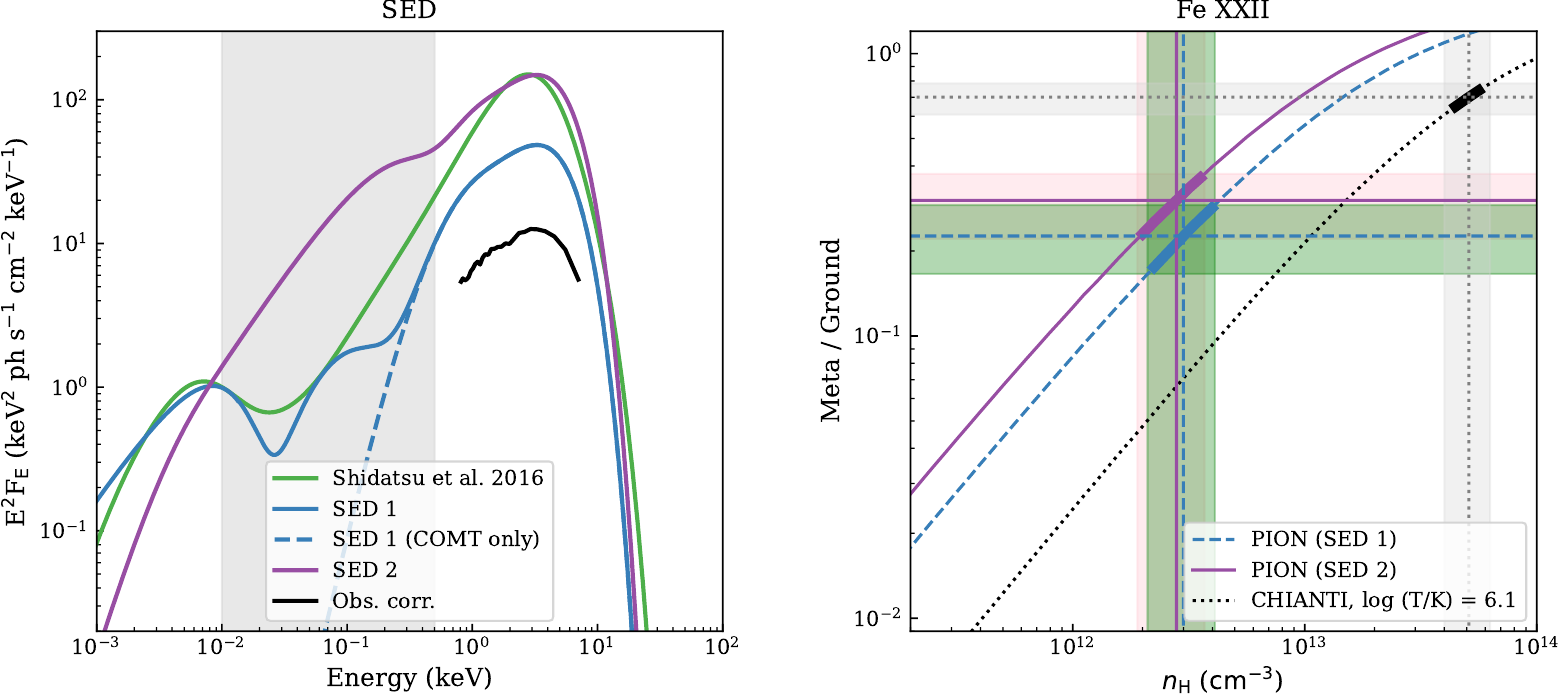}
    \caption{\textit{Left:} The spectral energy distributions SED-1 (blue) and SED-2 (purple) used in this work compared to that inferred by \citep{Shidatsu2016} (green). The black curve is the observed Chandra grating spectrum corrected for Galactic absorption and smoothed to show the continuum level. The shaded area in grey corresponds to $0.01-0.5$~keV where no observable data are available to constrain the SEDs. \textit{Right:} shows the metastable to ground level population ratios for Fe {\sc xxii} for SED-1 (blue) and SED-2 (purple) compared to that for collisional ionized equilibrium (black) processes alone. The latter was calculated using CHIANTI v10.0.1 \citep{DelZanna2021}. Increasing the intrinsic soft X-ray flux decreases the required density. It also increases the optical depth of the ground level transition at $\lambda11.77$ (${\rm 2s^2~2p~^2P_{1/2}}$). This means the line is more saturated, so its observed equivalent with does not increase linearly with the level population, so this also reduces the metastable ($\lambda11.92$: ${\rm 2s^2~2p~^2P_{3/2}}$) to ground level ratio from that given simply from the ratio of their observed equivalent widths and oscillator strengths.}
    \label{fig:plot_dcga_lpop}
\end{figure*}
We reanalyse the {\it Chandra}/HEG spectrum using
the photoionisation model {\tt pion} in {\sc spex} \citep[v3.05,][]{Kaastra2018}. 
Compared with the alternative photoionisation modelling using {\sc xstar} \citep{Kallman2001},
the advantage of {\tt pion} is that it can track the multiple 
radiative decay pathways which can populate the metastable levels \citep{Mao2017}. Additionally, it includes electron scattering \citep{Kastner1993} as well as atomic absorption in determining the observed continuum flux, and calculates on-the-fly the ion populations in spectral fitting 
self-consistently from the full SED model. 

We first explore the overall SED/absorber properties using
the {\it Chandra}/HEG data alone (black points in Fig.\ref{fig:plot_dcga_lpop}a). We use a single Comptonisation model as the continuum, with
two {\tt pion} absorbers 
partially covering the source \citep[see also,][]{Miller2008,Kallman2009}. 

Most of the observed absorption lines arise from the lower ionisation zone, which has a line of sight hydrogen column density of $\sim 5\times 10^{23}$~cm$^{-2}$, covering 60\% of the source, similar to previous analyses. 
The higher ionisation zone contributes most to the H- and He-like lines, especially Fe\,{\sc xxv} and Fe\,{\sc xxvi}. This zone 
covers most of the source, is faster, 
and 
has column density which is optically thick to electron scattering, such that the intrinsic luminosity at $1-10$~keV is a factor $\sim 3$ larger than the observed luminosity. Previous studies 
fit only to the ion columns, and did not include the effect of electron scattering in attenuating the illuminating flux. Thus the derived intrinsic SED (dashed blue line in Fig.\ref{fig:plot_dcga_lpop}a) is higher than observed (black) data.

The low ionisation zone produces the Fe {\sc xxii} $\lambda11.92$ metastable line. This is the first excited state above the ground, so can be populated by free electrons colliding with the ground state ion, raising the outer electron to this level. However, it can also be populated from above as a result of radiative cascades of electrons in higher excited levels. The upper levels
which can most efficiently radiatively decay down to the metastable level are those with the ${\rm 2s~2p^2}$ configuration. These are excited by photons with $h\nu\sim0.1$~keV. \cite{Mauche2003} says these are faint in their 
particular object, an accreting white dwarf, so that this process can be neglected, but our object is a black hole accretion disc, which is likely bright at these energies. 

We cannot directly observe the SED at $0.1$~keV due to interstellar absorption, which makes the grey shaded area in Fig.\ref{fig:plot_dcga_lpop}a unobservable. We cannot even uniquely know the $1-10$~keV X-ray luminosity due to electron scattering in the optically thick absorber. Hence we consider a range of different SED shapes and luminosities to span a 
plausible range of ionising continuua for our analysis. 
The green line in Fig.\ref{fig:plot_dcga_lpop}a is an 
SED derived by \citet{Shidatsu2016}, where the optical and UV flux match the observed level, but the X-rays are much brighter due to their even higher assumed electron scattering optical depth of $\tau\sim 3$. We add components to our Comptonisation continuum to make it resemble the shape of the SED of \citet{Shidatsu2016} below $0.1$~keV. This is shown as the solid blue line in Fig.\ref{fig:plot_dcga_lpop}a, SED-1. 

There could be much more flux at $0.1$~keV if the intrinsic luminosity derived by \citet{Shidatsu2016} is correct. We match 
the Comptonised continuum to their $1-10$~keV luminosity, and add the maximum disc continuum which is allowed by the 
contemporaneous UV data from {\it Swift}/UVOT$^1$. This SED-2 is shown as the purple line in Fig.\ref{fig:plot_dcga_lpop}a). Compared to SED-1, the $1-10$~keV intrinsic flux of SED-2 is higher, which is favoured by \citet{Neilsen2016, Shidatsu2016,Higginbottom2018}. More importantly for our study, SED-2 has more than an order of magnitude more flux at $0.1$~keV relative to the $1-10$~keV flux
 than SED-1. 
 


We fit the data with both SEDs. The high ionisation absorber should be closest to the source, so it filters the continuum seen by the lower ionisation absorber. The lower ionisation absorber (where the metastable lines are produced) is not optically thick, so the transmitted continuum which illuminates this zone must be around the level of the observed continuum. SED-1 gives results which are very similar to the initial fits with just a single Comptonisation component as the SED, with an optical depth of $\sim 2$ (column density of $\sim 2.5\times 10^{24}$~cm$^{-2}$) so that the observed luminosity is a factor $\sim 3$ smaller than the intrinsic when the covering fraction is taken into account. The higher luminosity of SED-2 requires a combination of a higher column density and a higher covering fraction in order to sufficiently attenuate the flux, but gives a similarly good fit to the data as the lines are highly saturated (Table~\ref{tbl:pars_sed}). 

The main density diagnostic line is from the ratio of the Fe {\sc xxii} metastable line (11.92 \AA) to the ground level resonance line (11.77 \AA). The black line in Figure~\ref{fig:plot_dcga_lpop}b shows the density derived from only collisional ionisation. In this solution, the ground state line is not strongly saturated, so the ratio of the observed equivalent width between the metastable and ground state directly estimates the ratio of level populations. 
The blue and magenta lines show the results of SED-1 and SED-2, respectively. Radiative cascades from excited states in both SEDs shift the derived density 
to $n_{\rm H}=(2.8\pm0.9)\times10^{12}~{\rm cm^{-3}}$,
down by over an order of magnitude from those reported previously \citep{Miller2008}. These SED's also shift the 
the inferred level population ratio down from that directly measured from the ratio of equivalent widths as 
the ground level line becomes optically thick (optical depth $\tau\sim 3.5-4$ in both SEDs).  

The decrease in density, together with the increased intrinsic luminosity, dramatically changes the inferred wind launch radius to be $\sim 10^{11}~\mathrm{cm}$.
By comparison, 
the Compton temperature for both SEDs is $\sim 6.4\times 10^6$~K, 
giving a Compton radius $R_{\rm IC}\sim 8.4\times10^{11}~{\rm cm}$.
Thermal winds can be launched from $0.1-0.2~R_{\rm IC}$ \citep{Woods1996}, clearly compatible with the new launch radius derived above.

\section{A specific thermal-radiative wind model}
\label{sec3}
\begin{figure}
    \centering
    \includegraphics[width=\hsize]{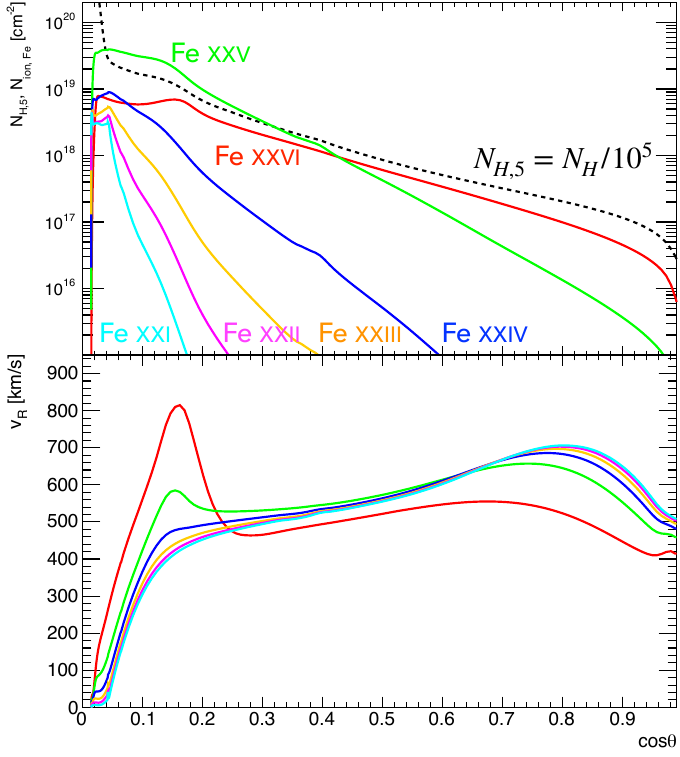}
    \caption{The inclination dependence of the hydrogen column density and  Fe ion columns (top) and average velocity weighted by those ion columns (bottom) for the thermal-radiative wind model. 
    Colours shows Fe\,{\sc xxvi} (red), {\sc xxv} (green), {\sc xxiv} (blue), {\sc xxiii} (yellow), {\sc xxii} (magenta), and {\sc xxi} (cyan).
     }
    \label{fig:integral}
\end{figure}
\begin{figure*}
    \centering
    \includegraphics[width=1.0\hsize]{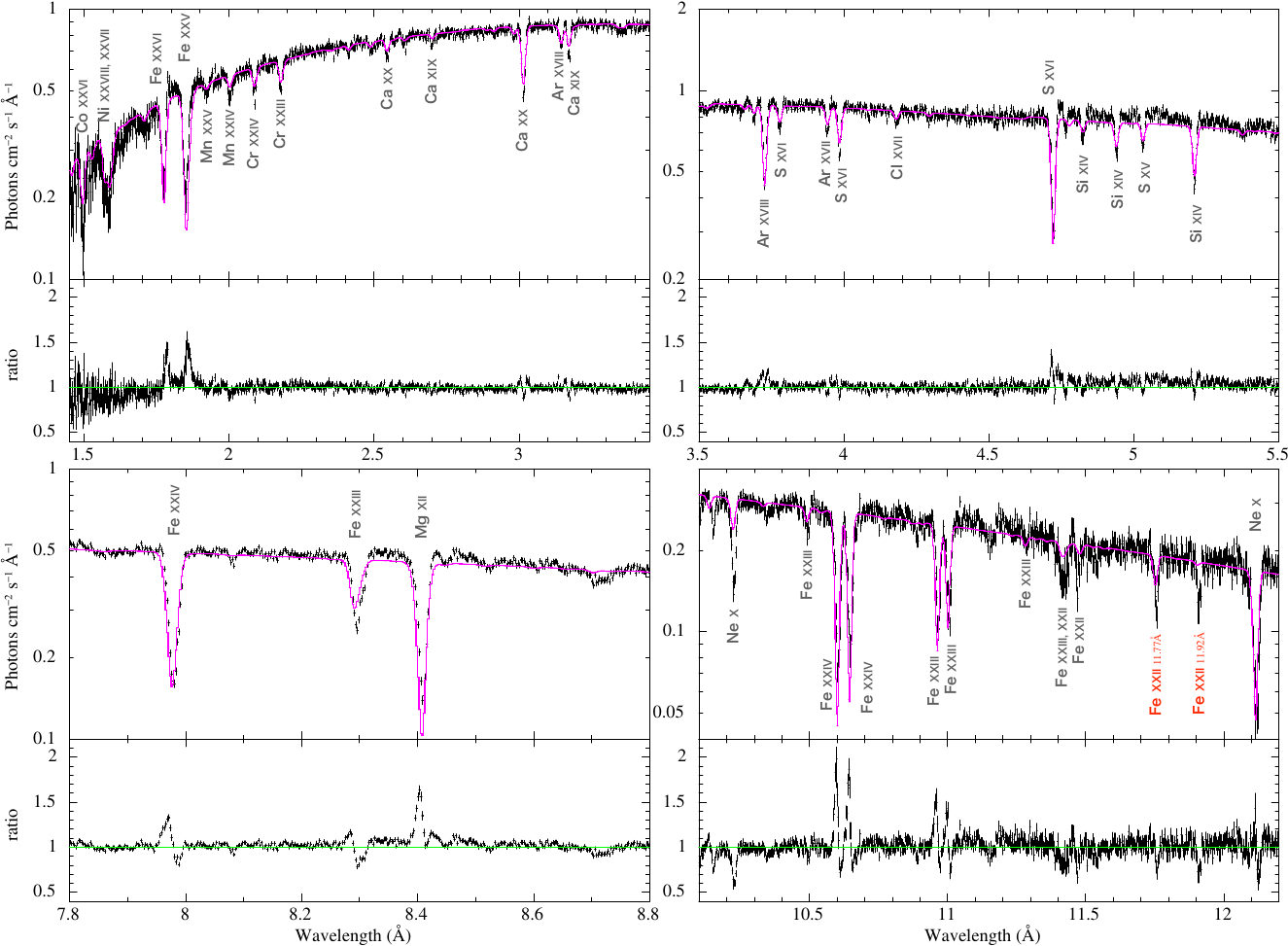}
    \caption{
    The {\it Chandra}/HEG spectrum of GRO J1655-40 (black) compared to the thermal-radiative wind model (magenta).
    We use the same energy range as \citet{Fukumura2017} for comparison easily.
    The model is calculated by radiation transfer via {\sc xstar} though the best fit line of sight.
    The continuum shape is the only additional free parameter, but its value is within a factor $\sim 4$ of the observed luminosity.  
    }
    \label{fig:global_thermal}
\end{figure*}
\begin{figure}
    \centering
    \includegraphics[width=1.0\hsize]{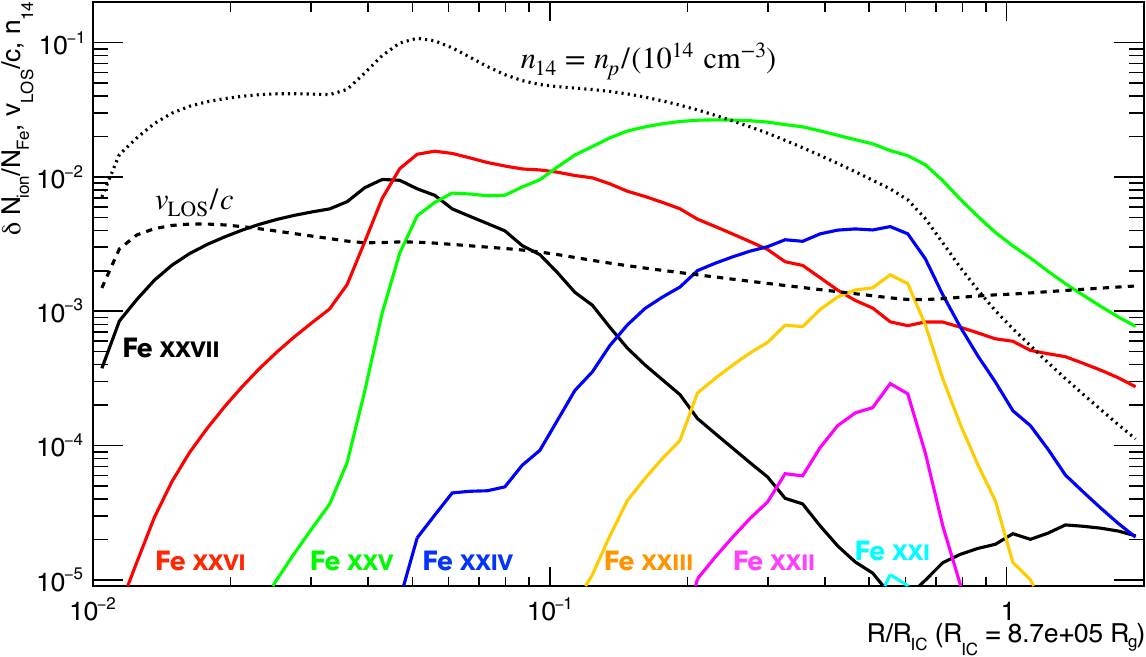}
    \caption{The radial properties of the optically thick thermal-radiative wind model. The hydrogen density (in units of $10^{14} ~\mathrm{cm^{-3}}$: grey dotted line) is almost constant above the disc, then drops sharply beyond the outer disc radius of $R= 4\times 10^{11} ~\mathrm{cm}$ ($0.5 R_\mathrm{IC}$). The local ion column densities divided by the total iron column density (solid coloured lines) show that the species are stratified, with 
    highly ionised ions produced at smaller radii and Fe\,{\sc xxii} peaking at the outer disc radius, where the density is $n=1.0\times 10^{12}~\mathrm{cm^{-3}}$.
    The radial velocity (in units of light speed: black dashed line), is also stratified, being $\sim 300$ km/s for 
    Fe\,{\sc xxii} and $\sim 500$ km/s for Fe\,{\sc xxvi}, close to that observed. 
    }
    \label{fig:radial_profile}
\end{figure}

We build a spectral model based on 
2-dimensional radiation hydrodynamic calculations of a 
thermal-radiative wind to compare to the observed spectrum by {\it Chandra}. 
We use the same radiation hydrodynamic simulation code as \citet{Tomaru2019, Tomaru2020b}. 
The code takes as input the broad band X-ray continuum from the central source, its luminosity and a disc size, and can calculate the dynamics of the gas ionised, heated, and accelerated by that radiation.
This radiation heating/cooling includes the line process (bound-bound), as well as bound-free, free-free, and Compton processes. The radiative acceleration includes bound-bound, bound-free and electron scattering.
These effects are pre-calculated by {\sc cloudy} \citep{Ferland2003} and converted to a look-up table as a function of ionisation parameter $\xi$ and temperature $T$ for the radiation hydrodynamic simulation.

The previous section has shown the difficulty in determining the intrinsic SED shape and luminosity. We take a single spectrum which is roughly the geometric mean of SED-1 and SED-2 in the X-ray bandpass (shown in the left panel in Fig. \ref{fig:sed_thermal})
with luminosity $L = 0.5 L_{\rm Edd}$.
The corresponding Compton temperature which is the equilibrium temperature at highest ionisation parameter is $T_{\rm IC} = 7.6\times 10^{6}~{\rm K}$.
The thermal equilibrium curve of photoionised gas irradiated by that SED is shown the right panel of Fig.\ref{fig:sed_thermal}.
Our simulation calculates the time evolution of gas on the disc surface which is illuminated by this SED. 

In our simulation, we use a time-dependent boundary condition on the disc surface, where the 
pressure ionisation parameter $\Xi =\Xi_c = L\exp(-\tau)/(4\pi R^2 c n kT)$ takes a constant critical value in order to set the base density of the wind. Here, $\tau$ is defined as the total optical depth of the wind and inner corona (atmosphere) along the line of sight from the centre to the disc surface, and $\Xi_c$ is set by the first thermal instability point, where the photoionised plasma makes a transition from a hot to cooler phase. 
The value of $\Xi_c$ depends on the shape of the irradiating spectrum (see Appendix).
We use the first instability point to set $\Xi_c$ since the cold gas becomes optically thick quickly due to line and bound-free opacity and the reprocessed emission is also important. 
However, we calculate this point assuming that the gas is optically thin, which means that the spectral shape is not different by that attenuation.
We will explore this effect in more detail using the realistic transmitted spectrum in a future work to show how different that is to the completely optically thin thermal winds with different boundary conditions \citep{Dyda2017, Luketic2010}.

Fig.\ref{fig:integral} shows the ion column densities and average velocity of each ion as a function of angle through the simulated wind structure (Fig.\ref{fig:2d_plot}). 
At high inclination angle, 
the hydrogen column density is optically thick with $N_{\rm H} \sim 10^{24}$~cm$^{-2}$.
At inclination angle $\theta =78-81^\circ (\cos \theta = 0.15-0.2) $, the Fe {\sc xxii} columns show the almost same value as required for the ground state 11.77\AA~line 
($N_{\rm xxii} = 1.9\pm 0.1 \times 10^{16}~{\rm cm^{-2}}$: \citealt{Miller2008}).
The velocity at these angles is also the same as observed.


We calculate transmitted spectra for the
high inclination angle region ( $75^\circ -85^\circ$ ) of this wind using the {\sc xstar} \citep{Kallman2001}. 
In this calculation, we do not include any artificial turbulent velocity to give additional broadening of the lines.
We  build a multiplicative tabular model ({\tt mtable}) in {\sc xspec} and fit the data  as ${\tt mtable}({\rm RHDwind})\times ({\tt ezdiskbb+nthcomp})$ with Galactic absorption.
The best fit parameter is $80^\circ$ and this transmitted spectrum matches fairly well to the observed data in terms of the ion states seen (Fig.\ref{fig:global_thermal}). One major exception is the 
absorption line from the metastable level of Fe\,{\sc xxii}, but this is due to our use of {\sc xstar}, which  does not include the radiation cascade populating this level. 
Additional mismatchs are in the Fe\,{\sc xxv} line, which may indicate there is significant emission from the wind in this line, and in the Ne\,{\sc x} line at 11.23\AA, 
which is probably due to  the rather low abundance of Neon assumed in {\sc xstar} 
(The default is $2.8\times 10^{-5}$ relative to hydrogen compared to e.g. $1.2\times 10^{-4}$ reported in \citealt{Lodders2009}, which is default abundance of {\sc spex}).

The density of our wind in the region where Fe\,{\sc xxii} is produced is similar (within a factor 2) to that required from the {\tt pion} analysis (see Fig.\ref{fig:radial_profile}). 
The overall optical depth to electron scattering along this sightline ($\cos \theta =0.15$) is $\tau_{\rm es}\sim 1.0$ (Fig.\ref{fig:integral}), 
a factor 2 lower than that required to attenuate the intrinsic luminosity down to the level observed.
Both these order 2 mismatches are likely due to the additional illumination of the outer disc by scattered flux in the wind which is not included in our radiative hydrodynamics code.
This will increase the density of the wind and hence decrease the observed flux.
Nonetheless, even without scattering, the thermal-radiative wind simulation shows a good overall match to the observed absorption structure. 
The wind density 
(dotted grey line in Fig. \ref{fig:radial_profile}) 
drops as $\sim R^{-1}$  over an order of magnitude change in radius, 
from $(0.1-1)R_{\rm out}=(0.05-0.5)R_{\rm IC}$.

This strongly contrasts with the expectations of optically thin thermal winds as $n\propto R^{-2}$ at high inclination angles \citep{Luketic2010}. 
These are the result of the base wind density of $n_0(R) \propto R^{-2}$ is set by the thermal instability via $\Xi_c$ with optically thin limit ($\tau=0$). 

 Hence the luminosity ionisation parameter $\xi=L/nR^2$ is constant, which is quite unlike the wind seen in these data.
 Instead, in our optically thick thermal-radiative wind, there is a wide range in ionisation parameter in the wind, with high ionisation material (Fe {\sc xxvii}: black and {\sc xxvi}: red) 
located at smaller radii while Fe\,{\sc xxii} (magenta) peaks towards the outer disc radius. This gives a higher line of sight velocity (dashed line) for the higher ionisation ion states, as observed. 

Thus the results of this single radiation hydrodynamic simulation of an optically thick 
thermal-radiative wind give a good overall match to all the observed features,
both the density derived from the metastable level of Fe {\sc xxii}
and the range of ion species seen, 
and to their observed velocities.
This shows that thermal-radiative winds are not excluded by 
this unique observation of GRO J1655-40, and in fact given that there no real free parameters barring the inclination, this proof of concept calculation gives strong evidence in favour of an optically thick, thermal-radiative wind as the explanation of these data. 
However, given the interest in magnetic winds in the literature, we revisit the magnetic wind solutions and compare the results of these models to our thermal-radiative wind solution. 

\section{The F17 magnetic wind model for \gro}

Current magnetic wind models use an ordered, large scale field threading the disc and rotating with it, and assume self-similar behaviour \citep{Fukumura2010, Chakravorty2013}.
In the models of F17 (based on previous work of \citealt{Contopoulos1994, Blandford1982}),
the field line geometry required is solved self consistently with the assumed power law radial density profile of the wind at its streamline base on the disc surface at radius $r_d$ such that 
$n(r_d)=n(r_0)(r_d/r_0)^{-q}$ where $r_0$ is the inner radius of the disc ($r_0 = 3.81\times 10^6~{\rm cm}$ for  \gro). 
In this approach, the wind density can be arbitrarily high.

F17 used an older version of {\sc xstar} (v2.2.1bn21), but this did not 
include the atomic data for the critical Fe\,{\sc xxii} lines
at 11.77~\AA\ and 11.92~\AA. 
We recompute the  magnetic wind model F17 designed to fit \gro using updated version of {\sc xstar} (v2.58e). 

%

\begin{figure*}
    \centering
    \includegraphics[width=\hsize]{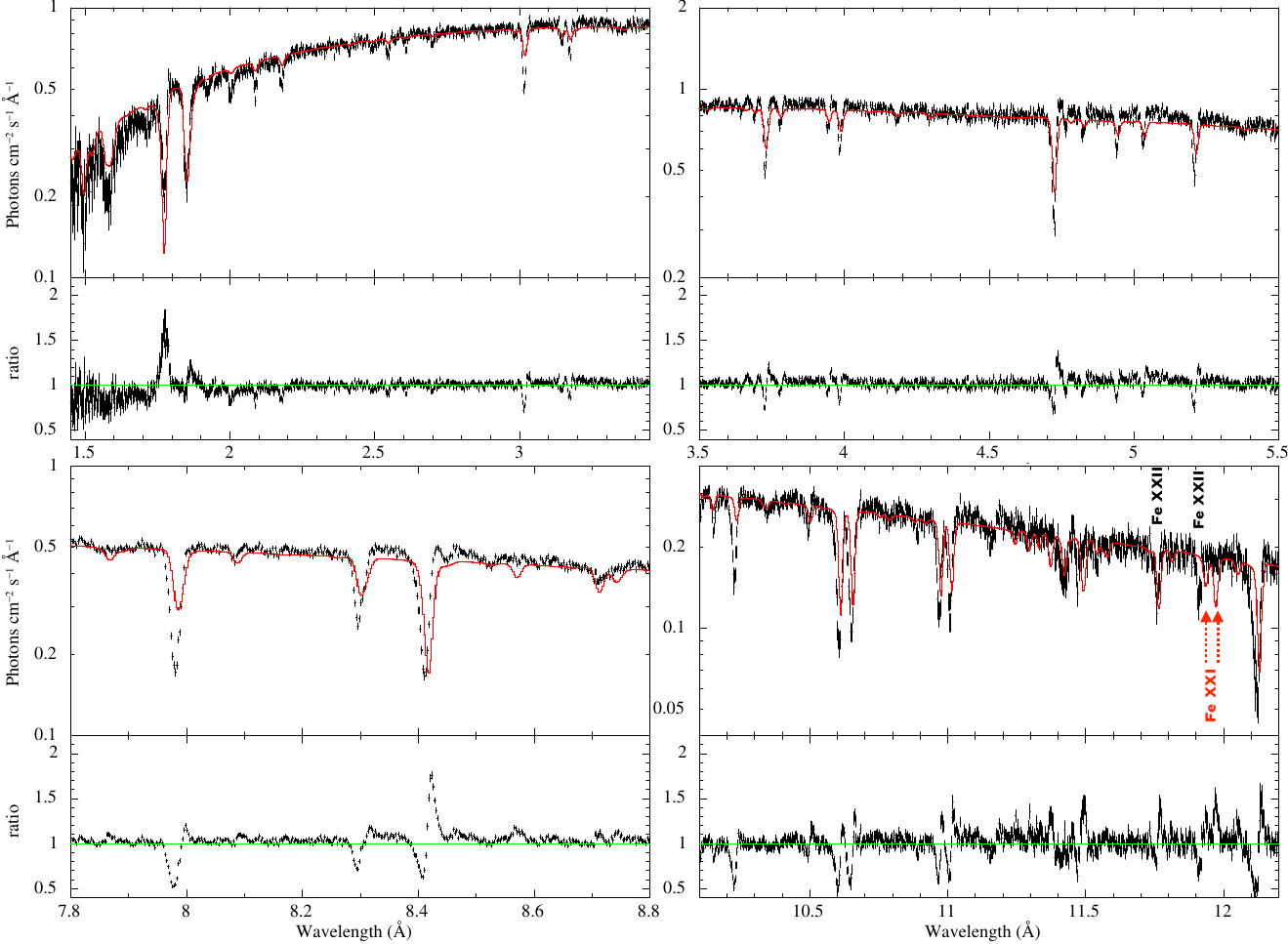}
    \caption{Our version of the MHD wind model of F17 i.e. with density $n(r_{sp},80^\circ)=1.4\times 10^{17} (r_{sp}/r_0)^{-1.2}$~cm$^{-3}$. The overall fit quality is similar to that shown in F17 across the range of ion states seen. }
    \label{fig:MHD_F17_1}
\end{figure*}

\begin{figure}
    \centering
    \includegraphics[width=\hsize]{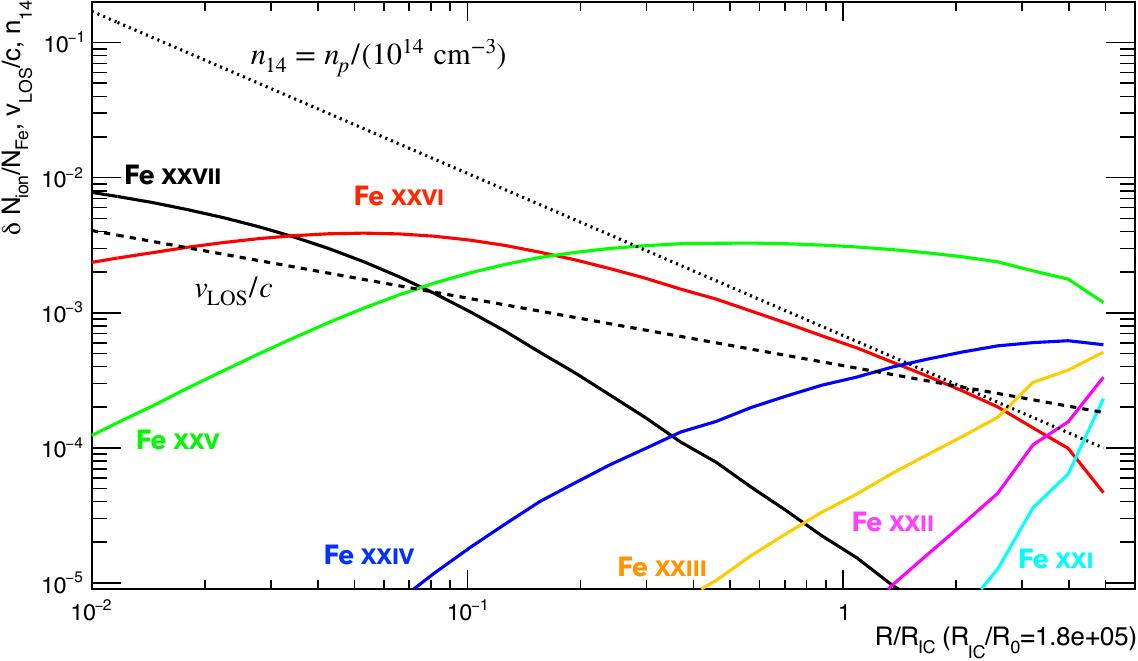}
    \caption{The radial dependence of the local fractional ion column along the $80^\circ$ line of sight  for our version of the magnetic wind model of F17 is shown in Fig. \ref{fig:MHD_F17_1}.
    Colours show Fe\,{\sc xxvii} (black), {\sc xxvi} (red),{\sc xxv} (green), {\sc xxiv} (blue), {\sc xxiii} (yellow), {\sc xxii} (magenta), and {\sc xxi} (cyan).  Fe\,{\sc xxii} and Fe\,{\sc xxi} are produced close to the outer radius of the wind.
    The dashed line shows the ratio of the radial velocity to the speed of light and the dotted line shows the hydrogen density normalised by $10^{14} {\rm cm^{-3}}$. 
    This figure only shows the outer region of the wind, where most of absorption lines are created. 
    }
    \label{fig:Bwind_radial}
\end{figure}

\begin{figure*}
    \centering
    \includegraphics[width=1.0\hsize]{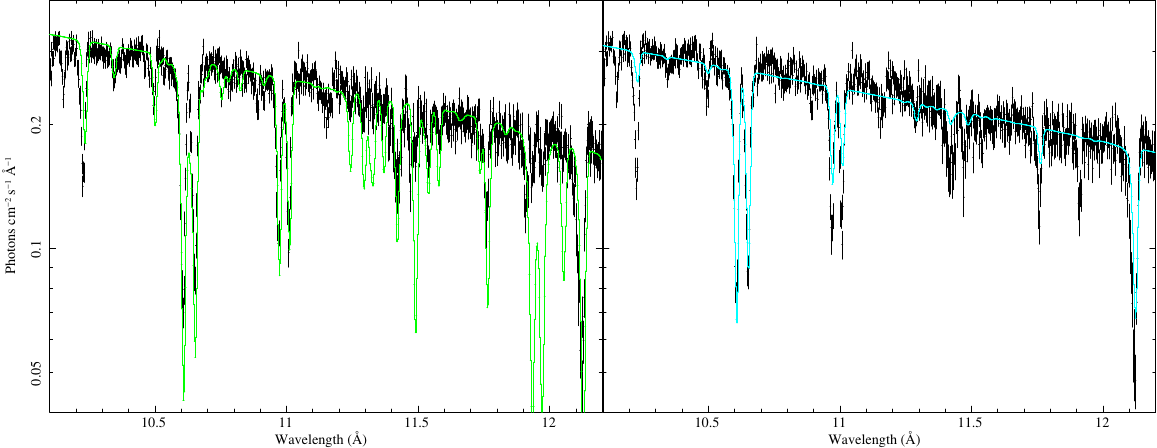}
    \caption{The same figure as F17 for the $10-12.2$~\AA\ region but different initial density $n = 3.0\times 10^{17} (R/R_0)^{-1.2} $ at $80^\circ$. Colours show different outer radius with $10^6 R_0$ (green) and $10^5 R_0$ (cyan).
    }
    \label{fig:MHD_F17_10p1_12p2A}
\end{figure*}

\begin{figure}
    \centering
    \includegraphics[width=\hsize]{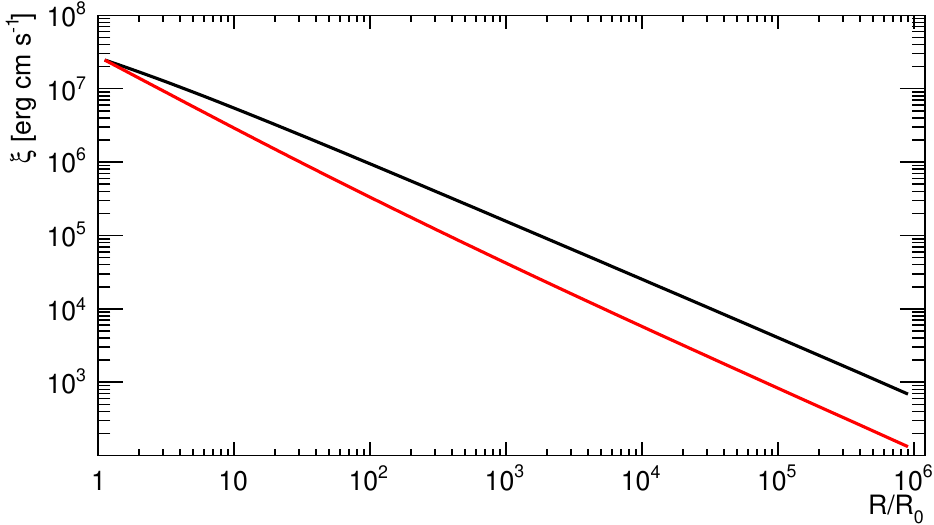}
    \caption{The radial distribution of ionisation parameter $\xi$ for the MHD wind model. Colours show optically thin limit (black) and with an attenuation factor $\exp (-\tau)$ (red).  }
    \label{fig:MHD_ionization}
\end{figure}

We calculate an MHD wind solution for these data
using the density profile of F17 where
$n(r, \theta) = n_0(r/r_0)^{-1.2}\exp(5(\theta-\pi/2))$.
F17 state that they use $n_0 = 9.3\times 10^{17},~\theta = 80^\circ$ i.e. $n(r, 80^\circ) = 3.9\times 10^{17} (r/r_0)^{-1.2}$ as the best fit for the broadband spectrum of {\it Chandra}/HEG but there were hidden overall scaling factors in their models. 
The actual density used is not easy to recover but may be around $n_0 = 3.5\times 10^{17}~\mathrm{cm^{-3}}$, so that the corresponding density profile at $\theta = 80^\circ$ is 
$n(r, \theta) = 1.4\times 10^{17} (r/r_0)^{-1.2}$ (Fukumura, private communication). 
The ionisation parameter is then $\xi = L/(nr^2) = 2.4\times 10^{7}(r/r_0)^{-0.8} \mathrm{erg~cm/s} $. 

We first calculate the absorption over the radial range of 
$r_0-10^6r_0$ at $80^\circ $ as used in F17,
but note that this gives an outer disc radius of $3.81\times 10^{12}$~cm, almost an order of magnitude larger than the expected outer radius of the 
disc in this system of $4.4\times 10^{11}$~cm. 
We also use the observed luminosity as the intrinsic luminosity as in F17 ($L= 5.0\times 10^{37} \mathrm{erg/s}$).
We split the column into 64 bins of equal logarithmic size in $\delta r/r$. 
The radial velocity along the line of sight for a streamline is $v_r(r, 80^\circ)/c = 0.17 (r/r_0)^{-0.5} \sim \cos 80^\circ (r/r_0)^{-0.5}$ i.e. the wind has slow radial outflow velocity in the equatorial plane (most of its velocity is in the azimuthal
direction, at the Keplerian velocity, $v_\phi(r)$), and is
radially accelerated by magneto-centrifugal forces.
Using these parameters, we solve for the radiation transfer along line of sight by chaining {\sc xstar}.
We calculate the transmitted spectra in the rest frame of the each wind element and the previous transmitted spectrum is used as next input spectrum.


Fig.\ref{fig:MHD_F17_1} shows the resulting fit to the {\it Chandra}/HEG data, where the model has additionally been Gaussian
smoothed to the instrument resolution using the {\tt gsmooth} 
model in {\sc xspec}.
We mark the positions of the Fe\,{\sc xxii} density diagnostic lines at 11.77~\AA\ and 11.92~\AA.
The older version of {\sc xstar} used by F17 did not include these lines so has no contribution from either of these lines. The newer {\sc xstar} version used in our calculation does include these absorption lines, but only includes collisional processes to populate the metastable level. The density in this MHD wind model is too small to make the metastable line by only colllisional processes despite the wind model being motivated by this feature. 

The absorption lines seen at 11.937~\AA\ and at 11.975~\AA\ are
instead from Fe\,{\sc xxi} (these transitions were again not 
included in the version of {\sc xstar} used in F17). 

Fig.\,\ref{fig:Bwind_radial} shows the radial structure of the wind along the $80^\circ$ of the outer region of the wind, where the most of absorption lines are created. 
The inner region is completely ionised so has no line absorption.
The wind only starts to become observable for 
$r/r_0 >  10^3$, where H-like iron (Fe\,{\sc xxvi}, red line) is able 
to form, followed by He-like (Fe\,{\sc xxv}, green line). 
The  Fe\,{\sc xxii} (B-like, magenta line) that produces the density diagnostic line 
peaks at the outer edge, where the ionisation parameter is $\xi= 2.4\times 10^7 (10^6)^{-0.8}\sim 500~{\rm erg ~cm~s^{-1}} $.
The density here is $\sim 10^{10}~{\rm cm^{-3}}$,
which is too low to form the metastable level even at the density suggested by our {\tt pion} analysis, let alone the density required for only collisional excitation. 

Increasing the density is not in itself a good solution.
We tried a model with 
$n(r, 80^\circ)=3\times 10^{17} (r/r_0)^{-1.2}$
but this dramatically increases the low ionisation species such as 
Fe\,{\sc xxi}, strongly overproducing the 11.937~\AA\ and 11.975~\AA\ absorption lines which are not in the data, as well as predicting multiple other 
unseen transitions from Fe\,{\sc xxii} and {\sc xvii}
to the right and left of the 
observed Fe\,{\sc xxii} lines at 11.427~\AA\ and 
11.491~\AA\ (left, Fig.\ref{fig:MHD_F17_10p1_12p2A}).
The observed lack of these additional Fe\,{\sc xxii}/{\sc xviii} lines (which were not included in the version of {\sc xstar} used by F17) clearly limit the contribution of the lower ionisation states of iron, and hence the amount of material with 
$\xi \gtrsim 500$. 
The ionisation parameter of this MHD wind monotonically decrease with radius, 
thus this lower limit of ionisation parameter must be at the maximum radius of the wind.

Using realistic outer radius $R_{\rm out} = 4.4\times 10^{11}~{\rm cm}$ as the maximum radius of the wind, which is almost 10 time smaller than that of F17, can give a wind with both higher ionisation parameter and higher density.
However, the density still lower than the observational value. 
We show the spectrum at $r= 10^5 r_0$ (right in Fig. \ref{fig:MHD_F17_10p1_12p2A}) with $n(r, 80^\circ) = 3\times 10^{17} (r/r_0)^{-1.2}$. 
This spectrum well describes the lower energy band without lines from Fe\,{\sc xxi}, but the density at $r= 10^5 r_0$ is $3\times 10^{11} {\rm cm^{-3}}$, which is still at least 10 times smaller than our derived density with the radiative cascades, and 100 times smaller than for collisional ionisation alone.
%
We consider what condition can  satisfy the observed 
density $n = 3\times 10^{12} {\rm cm^{-3}}$, the ionisation parameter $  500< \xi < 5000$ (this upper limit is taken from pion analysis),  and the observed luminosity $L=5\times 10^{37} {\rm erg/s}$ using the same density profile as $n\propto (r/r_0)^{-1.2}$. 
These equation give us the strong constrain of the radius where the Fe\,{\sc xxii} is dominant, and 
it is required to be $5.8 \times 10^{10} =1.5\times 10^{4} r_0< r_{\rm xxii} <1.8\times10^{11}~{\rm cm}= 4.7\times 10^4 r_0 $.
This also gives us a boundary condition on the density at $r=r_0$ of $3.1\times 10^{17} <n_{0,80} <12 \times 10^{17}$~cm$^{-3}$. 

%

Such a large density means that the model is not self consistent as this wind goes 
optically thick to electron scattering. This was not included in the models of F17, though the optical depth in their original model is also high, with $\tau_{\rm es} = \sigma_T \int^{r_{\rm out}}_{r_0} n(r, 80^\circ)dr = \sigma_T \int^{r_{\rm out}}_{r_0} 1.4\times 10^{17}(r/r_0)^{-1.2} dr =1.7$
Thus even original model F17 requires $\exp(1.7) =5.5$ times larger luminosity. 
Fig. \ref{fig:MHD_ionization} shows the effect of this attenuation on the ionisation parameter.
At the edge of the outer radius, the ionisation parameter is $5.5$ times smaller
so the ion calculations are not self consistent in F17. 
Our limits on the MHD inner wind density of $3\times 10^{17} <n_{0,80} < 12\times 10^{17}$, give corresponding optical depth of $3.4 < \tau_{\rm es}<13 $. This implies an extremely large intrinsic luminosity ($1.5\times 10^{39} < L < 2.2\times 10^{43}$) is required.

The only way(s) to recover a self-consistent solution 
with the observed density, ionisation parameter, and the observed luminosity with the radial dependence $n\propto (r/r_0)^{-1.2}$ is (are)
that the intrinsic source luminosity is much larger and/or 
that the wind only starts from some radius $r \gg r_0$.
For example, the column density in the observed (not completely ionised) ion species must be $\ge 5\times 10^{23}$~cm$^{-2}$,
corresponding to an optical depth of $\ge 0.3$. This requires that the wind only starts from a radius of $3\times 10^4r_0$ and that the intrinsic luminosity is $L_{\rm obs} \exp(\tau)=7\times 10^{37}$~ergs s$^{-1}$.
Alternatively, the wind has optical depth 1(3) if it extends down to $5\times 10^3r_0\ (225r_0)$, so the intrinsic luminosity is $\sim 10^{38} (10^{39})$~ergs s$^{-1}$.


Thus even the magnetic wind models have multiple issues with matching the data from \gro. 
The type of solutions proposed in F17 can give a good overall fit to the range of ion states seen for a line of sight density 
$n_{\rm LOS}(r)=1.4\times 10^{17}(r/r_0)^{-1.2}$, 
but this goes extremely optically
thick for a  wind extending from $r_0$ to $r_{out}$ as predicted for self-similar winds.

Thus even the MHD wind solutions require that the wind goes optically thick, so the source is intrinsically more luminous than observed. The previous MHD wind solutions did not include the effect of electron scattering on the ionising flux, so their ion populations are not self consistent. We show that including this means that
the MHD winds have great difficulty in matching the observed density diagnostic lines, which is ironic as this feature was the motivation for the magnetic wind solutions.

\section{Distinguishing magnetic and thermal wind models from line profiles in optically thin winds}
\begin{figure}
    \centering
    \includegraphics[width=\hsize]{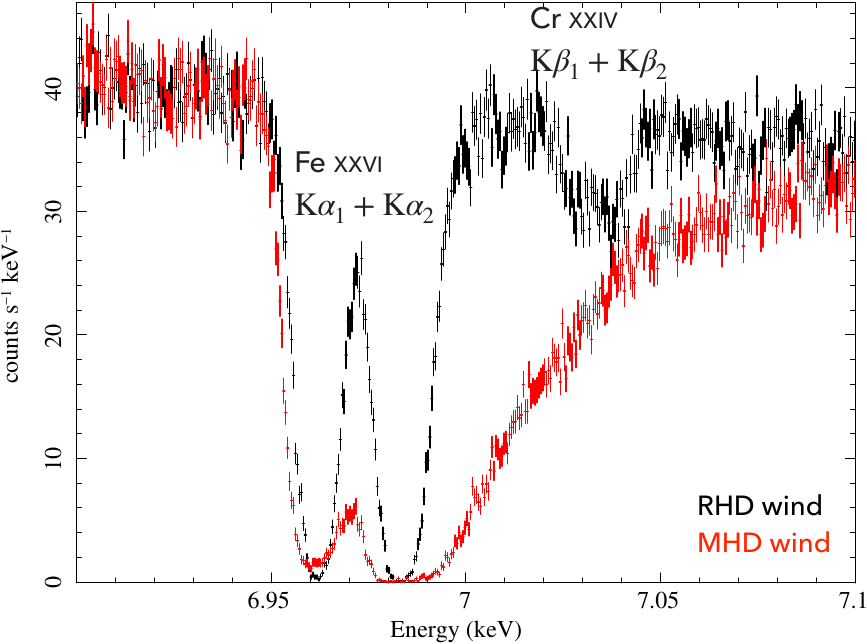}
    \caption{The simulation of observation by the microcalorimeter of {\it XRISM} with 30 ks for the thermal-radiative wind model (black) and the MHD wind model (red).   }
    \label{fig:xrism}
\end{figure}

We have shown for the first time that the thermal-radiative wind models can plausibly fit the iconic {\it Chandra}/HEG dataset for \gro, not just the broad range of ion species but even the specific metastable level, which was used to rule out the thermal wind models and motivate the MHD models. 
We also show for the first time that the MHD winds are not self-consistent in their neglect of electron scattering and cannot match the specific metastable level, which was their initial motivator.

Nonetheless, the thermal-radiative wind models also have some issues for these \gro\ data. They only predict a column density of $\tau \sim 1$, which is not enough to reduce the assumed X-ray flux down to that observed, and the best fit inclination angle is much 
higher than expected from the binary orbit. Plausibly, this is due to our neglect of flux scattering in the wind giving an extra irradiation source for the disc, but 
this is currently beyond the scope of our codes. 

Hence it could be argued that both models (MHD and thermal-radiative) 
give some mismatch with observations, so it is important to try to find a model independent way to distinguish the difference between them. 
Future X-ray spectrometers such as {\it XRISM}/Resolve and {\it Athena}/XIFU 
will give much higher resolution 
data showing the detailed line profile, so here we explore whether this can give
a clean diagnostic of the wind launching mechanism. 

Fig. \ref{fig:xrism} shows a
simulation of the expected {\it XRISM} spectra from the MHD (red) and thermal-radiative (black) wind models,
assuming the same exposure time as the that of {\it Chandra}/HEG (30 ks).
We pick the Fe\,{\sc xxvi} K$\alpha_{1,2}$ line region as this doublet is in a relatively clean part of the spectrum.

The MHD wind (red) shows a strong blue wing as in these models Fe\,{\sc xxvi} is produced over a large range in radii, and hence spans a large range in velocity as the MHD wind velocity is $\propto r^{-1/2}$. Higher velocity material also has higher velocity range in a given $d\log r$ segment, so the lines are broader as well as more blueshifted, so are less deep for constant column.
Thus the blue wing on MHD winds is a generic feature. 

By contrast, the thermal-radiative wind model gives an Fe\,{\sc xxvi} line which is relatively narrow and symmetric as the
wind velocity produced by thermal heating is relatively constant (Fig. \ref{fig:radial_profile}).
This constant velocity is a fairly generic feature of thermal winds, unless the wind launch radius is of order the disc size (\citealt{Tomaru2020}). The low velocity shear means that the resonance lines are very optically thick, so much weaker lines can contribute to the spectrum. The thermal-radiative wind predicts a noticeable contribution from the Cr\,{\sc xxiv} K$\beta_{1,2}$. Interestingly, the third order Chandra spectrum, which is the best resolution so far, has tentative evidence for a feature at this energy, though it was interpreted by \cite{Miller2015} as evidence for a separate velocity component in Fe\,{\sc xxvi} K$\alpha_{1,2}$.

Given that the wind in \gro\ appears to be more optically thick than our thermal-radiative wind simulation, we expect that the line profile in the \gro\ data would be more complex than that shown here. However, the line of sight through the wind in GX13+1 is probably comparable to this simulation (\citealt{Tomaru2020b}), so an observation by {\it XRISM}/Resolve of optically thin winds will distinguish between the models and reveal the wind driving mechanism.

\section{Summary}

We revisit the origin of the density sensitive Fe\,{\sc xxii} metastable line in 
the iconic Chandra grating spectrum of GRO J1655-40.
This was previously used to rule out an X-ray heated (thermal-radiative) wind in these data, which left  magnetic driving as the only viable mechanism. Since magnetic winds should scale with mass and mass accretion rate, this motivated 
unified models of magnetic winds which spanned both binaries and Quasars (F17). 
We reanalysed the data using a photo-ionisation code ({\sc pion} in the {\sc spex} package) which includes the contribution of radiative cascades as well as collisions in populating
the metastable level. This reduces the inferred wind density by more than an order of magnitude, removing the requirement for magnetic
winds. The derived column is also optically thick, so the source is intrinsically more luminous than observed. We use our state-of-the-art radiation hydrodynamics code to calculate the thermal-radiative wind expected in this source. Our specific simulation 
goes (marginally) optically thick, and gives a good overall match to all the observed
features, both the density derived from the metastable level of Fe\,{\sc xxii}, and the range of ion species seen, and their
observed velocities. This shows that thermal-radiative winds are not excluded
by this unique observation of GRO J1655-40.  In fact, given that there no
real free parameters barring the inclination, this proof of concept calculation
gives strong evidence in favour of an optically thick, thermal-radiative wind as
the explanation of these data, though we caution that such extreme winds are very hard to fully simulate.

There could still be magnetic winds in addition to the thermal-radiative wind, and a strong magnetic field might also suppress the thermal wind (\citealt{Waters2018}). Hence we also 
study the MHD wind used in F17 to fit these data. We show that this model
also implies that the wind is optically thick, so that the intrinsic luminosity is substantially larger than observed. This means the previous solutions of F17 were not self consistent as the effect of electron scattering reducing the illuminating flux throughout the wind was not included. 
We include this and show that unlike the thermal-radiative wind, the MHD wind requires additional free parameters of the wind size scale in order to match the ion species seen, even with our lower revised density for Fe\,{\sc xxii}.

Thus not only does the thermal-radiative wind give a good fit to \gro, the MHD wind does not. This not only removes the {\it requirement} for magnetic driving, but strongly {\it disfavours} it as being the origin of the wind seen in this unique 
observation, though we also show how future observations using high resolution calorimeter data from XRISM/Resolve and Athena/XIFU can easily distinguish thermal-radiative and MHD 
driving. 
This means there is no ``special" magnetic field structure, which has self-similarity across the entire mass scale as considered in previous work (F17).
Instead, it seems most likely that all the wind features seen in both the stellar and supermassive black holes can be explained by radiation (either from its energy which powers thermal winds via heating and/or its momentum powering UV line and dust driven winds). While the physics of these winds are complex, they can be directly calculated from the observed properties, unlike the magnetic winds which depend on an assumed  magnetic field configuration.
This provides motivation to tackle the complex astrophysics of radiative winds (e.g. \citealt{proga2004,nomura2017,doro2016,matthews2020,naddaf2021,ogawa2022} and references therein), to make 
quantitative models of AGN feedback for galaxy formation
(e.g \citealt{nomura2021,quera-bofarull2021}).

\section*{Acknowledgements}


This research has made use of data obtained from the Chandra Data Archive, and software provided by the Chandra X-ray Center (CXC).
CD and RT acknowledge support from the STFC consolidated grant ST/T000244/1.
RT acknowledges computer time on the Cray XC50 at the Center for Computational Astrophysics, National Observatory of Japan. SRON is supported financially by NWO, the Netherlands Organization for Scientific Research.
We thank M. Shidatsu for providing their SED data, and K. Fukumura for multiple discussions on the magnetic wind models. We also acknowledge useful discussions with T. Kallman, J. M. Miller, D. Rogantini, S. Zeegers, J. de Plaa, and J. S. Kaastra.

\section*{Data Availability}
The {\it Chandra} data are publicly available.
The {\tt pion} photoionisation code is publicly available as part of the {\sc spex} spectral fitting package.
{\sc xstar} code is also publicly available as part of the {\sc heasoft}, X-ray data analysis tool given by NASA.
{\sc cloudy} code is also publicly available.
 Access to the radiation hydrodynamic code is available on request from R.T. (ryota.tomaru@durham.ac.uk).

\bibliographystyle{mnras}
\bibliography{library} 

\begin{thebibliography}{}
\makeatletter
\relax
\def\mn@urlcharsother{\let\do\@makeother \do\$\do\&\do\#\do\^\do\_\do\%\do\~}
\def\mn@doi{\begingroup\mn@urlcharsother \@ifnextchar [ {\mn@doi@}
  {\mn@doi@[]}}
\def\mn@doi@[#1]#2{\def\@tempa{#1}\ifx\@tempa\@empty \href
  {http://dx.doi.org/#2} {doi:#2}\else \href {http://dx.doi.org/#2} {#1}\fi
  \endgroup}
\def\mn@eprint#1#2{\mn@eprint@#1:#2::\@nil}
\def\mn@eprint@arXiv#1{\href {http://arxiv.org/abs/#1} {{\tt arXiv:#1}}}
\def\mn@eprint@dblp#1{\href {http://dblp.uni-trier.de/rec/bibtex/#1.xml}
  {dblp:#1}}
\def\mn@eprint@#1:#2:#3:#4\@nil{\def\@tempa {#1}\def\@tempb {#2}\def\@tempc
  {#3}\ifx \@tempc \@empty \let \@tempc \@tempb \let \@tempb \@tempa \fi \ifx
  \@tempb \@empty \def\@tempb {arXiv}\fi \@ifundefined
  {mn@eprint@\@tempb}{\@tempb:\@tempc}{\expandafter \expandafter \csname
  mn@eprint@\@tempb\endcsname \expandafter{\@tempc}}}

\bibitem[\protect\citeauthoryear{{Abramowicz} \& {Klu{\'z}niak}}{{Abramowicz}
  \& {Klu{\'z}niak}}{2001}]{Abramowicz2001}
{Abramowicz} M.~A.,  {Klu{\'z}niak} W.,  2001, \mn@doi [\aap]
  {10.1051/0004-6361:20010791}, \href
  {https://ui.adsabs.harvard.edu/abs/2001A&A...374L..19A} {374, L19}

\bibitem[\protect\citeauthoryear{{Beer} \& {Podsiadlowski}}{{Beer} \&
  {Podsiadlowski}}{2002}]{Beer2002}
{Beer} M.~E.,  {Podsiadlowski} P.,  2002, \mn@doi [\mnras]
  {10.1046/j.1365-8711.2002.05189.x}, \href
  {https://ui.adsabs.harvard.edu/abs/2002MNRAS.331..351B} {331, 351}

\bibitem[\protect\citeauthoryear{Begelman \& McKee}{Begelman \&
  McKee}{1983}]{Begelman1983b}
Begelman M.~C.,  McKee C.~F.,  1983, \mn@doi [\apj] {10.1086/161179}, 271, 89

\bibitem[\protect\citeauthoryear{{Blandford} \& {Payne}}{{Blandford} \&
  {Payne}}{1982}]{Blandford1982}
{Blandford} R.~D.,  {Payne} D.~G.,  1982, \mn@doi [\mnras]
  {10.1093/mnras/199.4.883}, \href
  {http://adsabs.harvard.edu/abs/1982MNRAS.199..883B} {199, 883}

\bibitem[\protect\citeauthoryear{Chakravorty, Lee  \& Neilsen}{Chakravorty
  et~al.}{2013}]{Chakravorty2013}
Chakravorty S.,  Lee J.~C.,   Neilsen J.,  2013, \mn@doi [\mnras]
  {10.1093/mnras/stt1593}, 436, 560

\bibitem[\protect\citeauthoryear{{Contopoulos} \& {Lovelace}}{{Contopoulos} \&
  {Lovelace}}{1994}]{Contopoulos1994}
{Contopoulos} J.,  {Lovelace} R.~V.~E.,  1994, \mn@doi [\apj] {10.1086/174307},
  \href {https://ui.adsabs.harvard.edu/abs/1994ApJ...429..139C} {429, 139}

\bibitem[\protect\citeauthoryear{{Crenshaw}, {Kraemer}  \& {George}}{{Crenshaw}
  et~al.}{2003}]{Crenshaw2003}
{Crenshaw} D.~M.,  {Kraemer} S.~B.,   {George} I.~M.,  2003, \mn@doi [\araa]
  {10.1146/annurev.astro.41.082801.100328}, \href
  {https://ui.adsabs.harvard.edu/abs/2003ARA&A..41..117C} {41, 117}

\bibitem[\protect\citeauthoryear{{Cunningham}}{{Cunningham}}{1976}]{Cunningham1976}
{Cunningham} C.,  1976, \mn@doi [\apj] {10.1086/154636}, \href
  {https://ui.adsabs.harvard.edu/\#abs/1976ApJ...208..534C} {208, 534}

\bibitem[\protect\citeauthoryear{{Del Zanna}, {Dere}, {Young}  \& {Landi}}{{Del
  Zanna} et~al.}{2021}]{DelZanna2021}
{Del Zanna} G.,  {Dere} K.~P.,  {Young} P.~R.,   {Landi} E.,  2021, \mn@doi
  [\apj] {10.3847/1538-4357/abd8ce}, \href
  {https://ui.adsabs.harvard.edu/abs/2021ApJ...909...38D} {909, 38}

\bibitem[\protect\citeauthoryear{{D{\'{\i}}az Trigo} \& {Boirin}}{{D{\'{\i}}az
  Trigo} \& {Boirin}}{2016}]{DiazTrigo2016}
{D{\'{\i}}az Trigo} M.,  {Boirin} L.,  2016, \mn@doi [Astronomische
  Nachrichten] {10.1002/asna.201612315}, \href
  {http://adsabs.harvard.edu/abs/2016AN....337..368D} {337, 368}

\bibitem[\protect\citeauthoryear{{Done}, {Tomaru}  \& {Takahashi}}{{Done}
  et~al.}{2018}]{Done2018}
{Done} C.,  {Tomaru} R.,   {Takahashi} T.,  2018, \mn@doi [\mnras]
  {10.1093/mnras/stx2400}, \href
  {http://adsabs.harvard.edu/abs/2018MNRAS.473..838D} {473, 838}

\bibitem[\protect\citeauthoryear{{Dorodnitsyn}, {Kallman}  \&
  {Proga}}{{Dorodnitsyn} et~al.}{2016}]{doro2016}
{Dorodnitsyn} A.,  {Kallman} T.,   {Proga} D.,  2016, \mn@doi [\apj]
  {10.3847/0004-637X/819/2/115}, \href
  {https://ui.adsabs.harvard.edu/abs/2016ApJ...819..115D} {819, 115}

\bibitem[\protect\citeauthoryear{{Dyda}, {Dannen}, {Waters}  \& {Proga}}{{Dyda}
  et~al.}{2017}]{Dyda2017}
{Dyda} S.,  {Dannen} R.,  {Waters} T.,   {Proga} D.,  2017, \mn@doi [\mnras]
  {10.1093/mnras/stx406}, \href
  {https://ui.adsabs.harvard.edu/\#abs/2017MNRAS.467.4161D} {467, 4161}

\bibitem[\protect\citeauthoryear{Ferland}{Ferland}{2003}]{Ferland2003}
Ferland G.~J.,  2003, \mn@doi [\araa] {10.1146/annurev.astro.41.011802.094836},
  41, 517

\bibitem[\protect\citeauthoryear{{Fruscione} et~al.,}{{Fruscione}
  et~al.}{2006}]{fruscione2006}
{Fruscione} A.,  et~al., 2006, in {Silva} D.~R.,  {Doxsey} R.~E.,  eds,
  Society of Photo-Optical Instrumentation Engineers (SPIE) Conference Series
  Vol. 6270, Society of Photo-Optical Instrumentation Engineers (SPIE)
  Conference Series. p. 62701V, \mn@doi{10.1117/12.671760}

\bibitem[\protect\citeauthoryear{Fukumura, Kazanas, Contopoulos  \&
  Behar}{Fukumura et~al.}{2010}]{Fukumura2010}
Fukumura K.,  Kazanas D.,  Contopoulos I.,   Behar E.,  2010, \mn@doi [\apj]
  {10.1088/0004-637X/715/1/636}, 715, 636

\bibitem[\protect\citeauthoryear{{Fukumura}, {Kazanas}, {Shrader}, {Behar},
  {Tombesi}  \& {Contopoulos}}{{Fukumura} et~al.}{2017}]{Fukumura2017}
{Fukumura} K.,  {Kazanas} D.,  {Shrader} C.,  {Behar} E.,  {Tombesi} F.,
  {Contopoulos} I.,  2017, \mn@doi [Nature Astronomy]
  {10.1038/s41550-017-0062}, \href
  {http://adsabs.harvard.edu/abs/2017NatAs...1E..62F} {1, 0062}

\bibitem[\protect\citeauthoryear{{Greene}, {Bailyn}  \& {Orosz}}{{Greene}
  et~al.}{2001}]{Greene2001}
{Greene} J.,  {Bailyn} C.~D.,   {Orosz} J.~A.,  2001, \mn@doi [\apj]
  {10.1086/321411}, \href
  {https://ui.adsabs.harvard.edu/abs/2001ApJ...554.1290G} {554, 1290}

\bibitem[\protect\citeauthoryear{{Higginbottom}, {Proga}, {Knigge}, {Long},
  {Matthews}  \& {Sim}}{{Higginbottom} et~al.}{2014}]{Higginbottom2014}
{Higginbottom} N.,  {Proga} D.,  {Knigge} C.,  {Long} K.~S.,  {Matthews} J.~H.,
    {Sim} S.~A.,  2014, \mn@doi [\apj] {10.1088/0004-637X/789/1/19}, \href
  {https://ui.adsabs.harvard.edu/abs/2014ApJ...789...19H} {789, 19}

\bibitem[\protect\citeauthoryear{{Higginbottom}, {Knigge}, {Long}, {Matthews},
  {Sim}  \& {Hewitt}}{{Higginbottom} et~al.}{2018}]{Higginbottom2018}
{Higginbottom} N.,  {Knigge} C.,  {Long} K.~S.,  {Matthews} J.~H.,  {Sim}
  S.~A.,   {Hewitt} H.~A.,  2018, \mn@doi [\mnras] {10.1093/mnras/sty1599},
  \href {https://ui.adsabs.harvard.edu/\#abs/2018MNRAS.479.3651H} {479, 3651}

\bibitem[\protect\citeauthoryear{{Higginbottom}, {Knigge}, {Long}, {Matthews}
  \& {Parkinson}}{{Higginbottom} et~al.}{2019}]{Higginbottom2019}
{Higginbottom} N.,  {Knigge} C.,  {Long} K.~S.,  {Matthews} J.~H.,
  {Parkinson} E.~J.,  2019, \mn@doi [\mnras] {10.1093/mnras/stz310}, \href
  {https://ui.adsabs.harvard.edu/abs/2019MNRAS.484.4635H} {484, 4635}

\bibitem[\protect\citeauthoryear{{Higginbottom}, {Knigge}, {Sim}, {Long},
  {Matthews}, {Hewitt}, {Parkinson}  \& {Mangham}}{{Higginbottom}
  et~al.}{2020}]{Higginbottom2020}
{Higginbottom} N.,  {Knigge} C.,  {Sim} S.~A.,  {Long} K.~S.,  {Matthews}
  J.~H.,  {Hewitt} H.~A.,  {Parkinson} E.~J.,   {Mangham} S.~W.,  2020, \mn@doi
  [\mnras] {10.1093/mnras/staa209}, \href
  {https://ui.adsabs.harvard.edu/abs/2020MNRAS.492.5271H} {492, 5271}

\bibitem[\protect\citeauthoryear{{Hjellming} \& {Rupen}}{{Hjellming} \&
  {Rupen}}{1995}]{Hjellming1995}
{Hjellming} R.~M.,  {Rupen} M.~P.,  1995, \mn@doi [\nat] {10.1038/375464a0},
  \href {https://ui.adsabs.harvard.edu/abs/1995Natur.375..464H} {375, 464}

\bibitem[\protect\citeauthoryear{{Huenemoerder} et~al.,}{{Huenemoerder}
  et~al.}{2011}]{huenemoerder2011}
{Huenemoerder} D.~P.,  et~al., 2011, \mn@doi [\aj]
  {10.1088/0004-6256/141/4/129}, \href
  {https://ui.adsabs.harvard.edu/abs/2011AJ....141..129H} {141, 129}

\bibitem[\protect\citeauthoryear{{Kaastra}}{{Kaastra}}{2017}]{Kaastra2017}
{Kaastra} J.~S.,  2017, \mn@doi [\aap] {10.1051/0004-6361/201629319}, \href
  {https://ui.adsabs.harvard.edu/abs/2017A&A...605A..51K} {605, A51}

\bibitem[\protect\citeauthoryear{{Kaastra}, {Raassen}, {de Plaa}  \&
  {Gu}}{{Kaastra} et~al.}{2018}]{Kaastra2018}
{Kaastra} J.~S.,  {Raassen} A.~J.~J.,  {de Plaa} J.,   {Gu} L.,  2018, {SPEX
  X-ray spectral fitting package}, \mn@doi{10.5281/zenodo.2419563}

\bibitem[\protect\citeauthoryear{{Kallman} \& {Bautista}}{{Kallman} \&
  {Bautista}}{2001}]{Kallman2001}
{Kallman} T.,  {Bautista} M.,  2001, \mn@doi [\apjs] {10.1086/319184}, \href
  {http://adsabs.harvard.edu/abs/2001ApJS..133..221K} {133, 221}

\bibitem[\protect\citeauthoryear{{Kallman}, {Bautista}, {Goriely}, {Mendoza},
  {Miller}, {Palmeri}, {Quinet}  \& {Raymond}}{{Kallman}
  et~al.}{2009}]{Kallman2009}
{Kallman} T.~R.,  {Bautista} M.~A.,  {Goriely} S.,  {Mendoza} C.,  {Miller}
  J.~M.,  {Palmeri} P.,  {Quinet} P.,   {Raymond} J.,  2009, \mn@doi [\apj]
  {10.1088/0004-637X/701/2/865}, \href
  {https://ui.adsabs.harvard.edu/\#abs/2009ApJ...701..865K} {701, 865}

\bibitem[\protect\citeauthoryear{{Kastner}}{{Kastner}}{1993}]{Kastner1993}
{Kastner} S.~O.,  1993, \mn@doi [\ssr] {10.1007/BF00754512}, \href
  {https://ui.adsabs.harvard.edu/abs/1993SSRv...65..317K} {65, 317}

\bibitem[\protect\citeauthoryear{{Kimura} \& {Done}}{{Kimura} \&
  {Done}}{2019}]{Kimura2019}
{Kimura} M.,  {Done} C.,  2019, \mn@doi [\mnras] {10.1093/mnras/sty2736}, \href
  {https://ui.adsabs.harvard.edu/\#abs/2019MNRAS.482..626K} {482, 626}

\bibitem[\protect\citeauthoryear{{Laha}, {Reynolds}, {Reeves}, {Kriss},
  {Guainazzi}, {Smith}, {Veilleux}  \& {Proga}}{{Laha} et~al.}{2021}]{Laha2021}
{Laha} S.,  {Reynolds} C.~S.,  {Reeves} J.,  {Kriss} G.,  {Guainazzi} M.,
  {Smith} R.,  {Veilleux} S.,   {Proga} D.,  2021, \mn@doi [Nature Astronomy]
  {10.1038/s41550-020-01255-2}, \href
  {https://ui.adsabs.harvard.edu/abs/2021NatAs...5...13L} {5, 13}

\bibitem[\protect\citeauthoryear{{Lodders}, {Palme}  \& {Gail}}{{Lodders}
  et~al.}{2009}]{Lodders2009}
{Lodders} K.,  {Palme} H.,   {Gail} H.~P.,  2009, \mn@doi [Landolt
  B\&ouml;rnstein] {10.1007/978-3-540-88055-4_34}, \href
  {https://ui.adsabs.harvard.edu/abs/2009LanB...4B..712L} {4B, 712}

\bibitem[\protect\citeauthoryear{Luketic, Proga, Kallman, Raymond  \&
  Miller}{Luketic et~al.}{2010}]{Luketic2010}
Luketic S.,  Proga D.,  Kallman T.~R.,  Raymond J.~C.,   Miller J.~M.,  2010,
  \mn@doi [\apj] {10.1088/0004-637X/719/1/515}, 719, 515

\bibitem[\protect\citeauthoryear{{Mao}, {Kaastra}, {Mehdipour}, {Raassen}, {Gu}
   \& {Miller}}{{Mao} et~al.}{2017}]{Mao2017}
{Mao} J.,  {Kaastra} J.~S.,  {Mehdipour} M.,  {Raassen} A.~J.~J.,  {Gu} L.,
  {Miller} J.~M.,  2017, \mn@doi [\aap] {10.1051/0004-6361/201731378}, \href
  {https://ui.adsabs.harvard.edu/abs/2017A&A...607A.100M} {607, A100}

\bibitem[\protect\citeauthoryear{{Matthews}, {Knigge}, {Higginbottom}, {Long},
  {Sim}, {Mangham}, {Parkinson}  \& {Hewitt}}{{Matthews}
  et~al.}{2020}]{matthews2020}
{Matthews} J.~H.,  {Knigge} C.,  {Higginbottom} N.,  {Long} K.~S.,  {Sim}
  S.~A.,  {Mangham} S.~W.,  {Parkinson} E.~J.,   {Hewitt} H.~A.,  2020, \mn@doi
  [\mnras] {10.1093/mnras/staa136}, \href
  {https://ui.adsabs.harvard.edu/abs/2020MNRAS.492.5540M} {492, 5540}

\bibitem[\protect\citeauthoryear{{Mauche}, {Liedahl}  \& {Fournier}}{{Mauche}
  et~al.}{2003}]{Mauche2003}
{Mauche} C.~W.,  {Liedahl} D.~A.,   {Fournier} K.~B.,  2003, \mn@doi [\apjl]
  {10.1086/375684}, \href
  {https://ui.adsabs.harvard.edu/abs/2003ApJ...588L.101M} {588, L101}

\bibitem[\protect\citeauthoryear{{Mauche}, {Liedahl}  \& {Fournier}}{{Mauche}
  et~al.}{2004}]{Mauche2004}
{Mauche} C.~W.,  {Liedahl} D.~A.,   {Fournier} K.~B.,  2004, in {Vrielmann} S.,
   {Cropper} M.,  eds,  Astronomical Society of the Pacific Conference Series
  Vol. 315, IAU Colloq. 190: Magnetic Cataclysmic Variables. p.~124 (\mn@eprint
  {arXiv} {astro-ph/0301633})

\bibitem[\protect\citeauthoryear{{Mehdipour}, {Kaastra}  \&
  {Kallman}}{{Mehdipour} et~al.}{2016}]{Mehdipour2016}
{Mehdipour} M.,  {Kaastra} J.~S.,   {Kallman} T.,  2016, \mn@doi [\aap]
  {10.1051/0004-6361/201628721}, \href
  {https://ui.adsabs.harvard.edu/abs/2016A&A...596A..65M} {596, A65}

\bibitem[\protect\citeauthoryear{Miller, Raymond, Fabian, Steeghs, Homan,
  Reynolds, van~der Klis  \& Wijnands}{Miller et~al.}{2006a}]{Miller2006}
Miller J.~M.,  Raymond J.,  Fabian A.,  Steeghs D.,  Homan J.,  Reynolds C.~S.,
   van~der Klis M.,   Wijnands R.,  2006a, \mn@doi [Nature]
  {10.1038/nature04912}, 441, 953

\bibitem[\protect\citeauthoryear{{Miller} et~al.,}{{Miller}
  et~al.}{2006b}]{Miller2006b}
{Miller} J.~M.,  et~al., 2006b, \mn@doi [\apj] {10.1086/504673}, \href
  {https://ui.adsabs.harvard.edu/#abs/2006ApJ...646..394M} {646, 394}

\bibitem[\protect\citeauthoryear{Miller, Raymond, Reynolds, Fabian, Kallman  \&
  Homan}{Miller et~al.}{2008}]{Miller2008}
Miller J.~M.,  Raymond J.,  Reynolds C.~S.,  Fabian a.~C.,  Kallman T.~R.,
  Homan J.,  2008, \mn@doi [Astrophys. J] {10.1086/588521}, 680, 1359

\bibitem[\protect\citeauthoryear{Miller, Fabian, Kaastra, Kallman, King, Proga,
  Raymond  \& Reynolds}{Miller et~al.}{2015}]{Miller2015}
Miller J.~M.,  Fabian A.~C.,  Kaastra J.,  Kallman T.,  King A.~L.,  Proga D.,
  Raymond J.,   Reynolds C.~S.,  2015, \mn@doi [\apj]
  {10.1088/0004-637X/814/2/87}, 814, 87

\bibitem[\protect\citeauthoryear{Miller et~al.,}{Miller
  et~al.}{2016}]{Miller2016}
Miller J.~M.,  et~al., 2016, \mn@doi [\apj] {10.3847/2041-8205/821/1/L9}, 821,
  L9

\bibitem[\protect\citeauthoryear{{Naddaf}, {Czerny}  \& {Szczerba}}{{Naddaf}
  et~al.}{2021}]{naddaf2021}
{Naddaf} M.-H.,  {Czerny} B.,   {Szczerba} R.,  2021, \mn@doi [\apj]
  {10.3847/1538-4357/ac139d}, \href
  {https://ui.adsabs.harvard.edu/abs/2021ApJ...920...30N} {920, 30}

\bibitem[\protect\citeauthoryear{{Neilsen}, {Rahoui}, {Homan}  \&
  {Buxton}}{{Neilsen} et~al.}{2016}]{Neilsen2016}
{Neilsen} J.,  {Rahoui} F.,  {Homan} J.,   {Buxton} M.,  2016, \mn@doi [\apj]
  {10.3847/0004-637X/822/1/20}, \href
  {http://adsabs.harvard.edu/abs/2016ApJ...822...20N} {822, 20}

\bibitem[\protect\citeauthoryear{{Nomura} \& {Ohsuga}}{{Nomura} \&
  {Ohsuga}}{2017}]{nomura2017}
{Nomura} M.,  {Ohsuga} K.,  2017, \mn@doi [\mnras] {10.1093/mnras/stw2877},
  \href {https://ui.adsabs.harvard.edu/#abs/2017MNRAS.465.2873N} {465, 2873}

\bibitem[\protect\citeauthoryear{{Nomura}, {Omukai}  \& {Ohsuga}}{{Nomura}
  et~al.}{2021}]{nomura2021}
{Nomura} M.,  {Omukai} K.,   {Ohsuga} K.,  2021, \mn@doi [\mnras]
  {10.1093/mnras/stab2214}, \href
  {https://ui.adsabs.harvard.edu/abs/2021MNRAS.507..904N} {507, 904}

\bibitem[\protect\citeauthoryear{{Ogawa}, {Ueda}, {Wada}  \&
  {Mizumoto}}{{Ogawa} et~al.}{2022}]{ogawa2022}
{Ogawa} S.,  {Ueda} Y.,  {Wada} K.,   {Mizumoto} M.,  2022, \mn@doi [\apj]
  {10.3847/1538-4357/ac3cb9}, \href
  {https://ui.adsabs.harvard.edu/abs/2022ApJ...925...55O} {925, 55}

\bibitem[\protect\citeauthoryear{{Orosz}, {Remillard}, {Bailyn}  \&
  {McClintock}}{{Orosz} et~al.}{1997}]{Orosz1997}
{Orosz} J.~A.,  {Remillard} R.~A.,  {Bailyn} C.~D.,   {McClintock} J.~E.,
  1997, \mn@doi [\apjl] {10.1086/310553}, \href
  {https://ui.adsabs.harvard.edu/abs/1997ApJ...478L..83O} {478, L83}

\bibitem[\protect\citeauthoryear{Ponti, Fender, Begelman, Dunn, Neilsen  \&
  Coriat}{Ponti et~al.}{2012}]{ponti2012}
Ponti G.,  Fender R.~P.,  Begelman M.~C.,  Dunn R. J.~H.,  Neilsen J.,   Coriat
  M.,  2012, \mn@doi [\mnras: Letters] {10.1111/j.1745-3933.2012.01224.x}, 422,
  11

\bibitem[\protect\citeauthoryear{Proga \& Kallman}{Proga \&
  Kallman}{2002}]{Proga2002}
Proga D.,  Kallman T.~R.,  2002, \mn@doi [\apj] {10.1086/324534}, 20, 455

\bibitem[\protect\citeauthoryear{Proga \& Kallman}{Proga \&
  Kallman}{2004}]{proga2004}
Proga D.,  Kallman T.~R.,  2004, \mn@doi [\apj] {10.1086/425117}, 616, 688

\bibitem[\protect\citeauthoryear{{Quera-Bofarull}, {Done}, {Lacey}, {Nomura}
  \& {Ohsuga}}{{Quera-Bofarull} et~al.}{2021}]{quera-bofarull2021}
{Quera-Bofarull} A.,  {Done} C.,  {Lacey} C.~G.,  {Nomura} M.,   {Ohsuga} K.,
  2021, arXiv e-prints, \href
  {https://ui.adsabs.harvard.edu/abs/2021arXiv211102742Q} {p. arXiv:2111.02742}

\bibitem[\protect\citeauthoryear{{Shidatsu}, {Done}  \& {Ueda}}{{Shidatsu}
  et~al.}{2016}]{Shidatsu2016}
{Shidatsu} M.,  {Done} C.,   {Ueda} Y.,  2016, \mn@doi [\apj]
  {10.3847/0004-637X/823/2/159}, \href
  {http://adsabs.harvard.edu/abs/2016ApJ...823..159S} {823, 159}

\bibitem[\protect\citeauthoryear{{Tomaru}, {Done}, {Ohsuga}, {Nomura}  \&
  {Takahashi}}{{Tomaru} et~al.}{2019}]{Tomaru2019}
{Tomaru} R.,  {Done} C.,  {Ohsuga} K.,  {Nomura} M.,   {Takahashi} T.,  2019,
  \mn@doi [\mnras] {10.1093/mnras/stz2738}, \href
  {https://ui.adsabs.harvard.edu/abs/2019MNRAS.490.3098T} {490, 3098}

\bibitem[\protect\citeauthoryear{{Tomaru}, {Done}, {Ohsuga}, {Odaka}  \&
  {Takahashi}}{{Tomaru} et~al.}{2020a}]{Tomaru2020}
{Tomaru} R.,  {Done} C.,  {Ohsuga} K.,  {Odaka} H.,   {Takahashi} T.,  2020a,
  \mn@doi [\mnras] {10.1093/mnras/staa961}, \href
  {https://ui.adsabs.harvard.edu/abs/2020MNRAS.494.3413T} {494, 3413}

\bibitem[\protect\citeauthoryear{{Tomaru}, {Done}, {Ohsuga}, {Odaka}  \&
  {Takahashi}}{{Tomaru} et~al.}{2020b}]{Tomaru2020b}
{Tomaru} R.,  {Done} C.,  {Ohsuga} K.,  {Odaka} H.,   {Takahashi} T.,  2020b,
  \mn@doi [\mnras] {10.1093/mnras/staa2254}, \href
  {https://ui.adsabs.harvard.edu/abs/2020MNRAS.497.4970T} {497, 4970}

\bibitem[\protect\citeauthoryear{Ueda, Murakami, Yamaoka, Dotani  \&
  Ebisawa}{Ueda et~al.}{2004}]{Ueda2004}
Ueda Y.,  Murakami H.,  Yamaoka K.,  Dotani T.,   Ebisawa K.,  2004, \mn@doi
  [\apj] {10.1086/420973}, 609, 325

\bibitem[\protect\citeauthoryear{{Waters} \& {Proga}}{{Waters} \&
  {Proga}}{2018}]{Waters2018}
{Waters} T.,  {Proga} D.,  2018, \mn@doi [\mnras] {10.1093/mnras/sty2398},
  \href {https://ui.adsabs.harvard.edu/abs/2018MNRAS.481.2628W} {481, 2628}

\bibitem[\protect\citeauthoryear{{Woods}, {Klein}, {Castor}, {McKee}  \&
  {Bell}}{{Woods} et~al.}{1996}]{Woods1996}
{Woods} D.~T.,  {Klein} R.~I.,  {Castor} J.~I.,  {McKee} C.~F.,   {Bell} J.~B.,
   1996, \mn@doi [\apj] {10.1086/177101}, \href
  {https://ui.adsabs.harvard.edu/#abs/1996ApJ...461..767W} {461, 767}

\makeatother
\end{thebibliography}


\newpage
\appendix
\section{Radiation hydrodynamic simulation}

\begin{figure}
    \centering
    \includegraphics[width=\hsize]{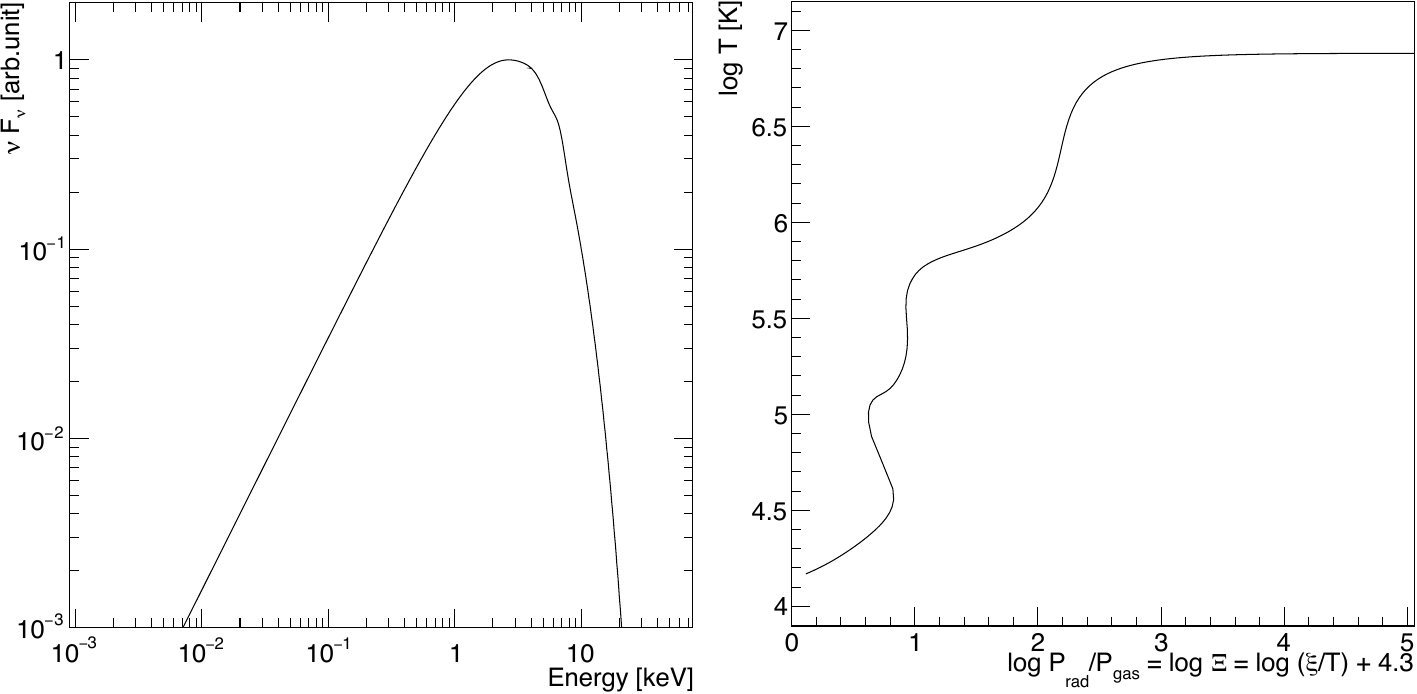}
    \caption{The broadband X-ray continuum (left) and thermal equilibrium curve of photoionised plasma irradiated by that (right). }
    \label{fig:sed_thermal}
\end{figure}

We calculate density, velocity and temperature distribution of the wind 
in a 2D (axisymmetric) radiation hydrodynamic simulation, 
which is  the same code as \citet{Tomaru2019, Tomaru2020b}, but we review it here for completeness. 
The code requires the SED, its luminosity and the disc outer radius as input parameters.
We take the SED  from the simultaneous {\it RXTE} observation (OBSID:91702-01-19-00) (left in Fig.\ref{fig:sed_thermal}). 
Using photoionisation code {\sc cloudy}, we also calculate the thermal equilibrium curve of photo-ionised plasma irradiated by that SED (right in Fig. \ref{fig:sed_thermal}).

Using {\sc cloudy} and this SED as the input parameter, we calculate radiation heating/cooling and the force multiplier, which is the ratio of effective optical depth to that of electron scattering as functions of $\xi$ and $T$ in optically thin limit. 
The radiative heating/cooling includes Compton process, free-free, bound-free, free-bound, bound-bound, and force multiplier includes bound-bound and bound-free. 
We run radiation hydrodynamic simulation by solving the equations in \citet{Tomaru2019}.
We use spherical polar coordinates, and
set the computational domain in R direction as $R_{\rm in}=0.01 R_{\rm IC}$
and $R_{\rm out} = 2 R_{\rm IC}$.
The radial grid spacing is logarithmic, calculated by 
\begin{equation}
R_i = R_\mathrm{\mathrm{in}} (R_\mathrm{out}/R_\mathrm{in})^{i/N_R}, ~ \text{($0\leq i \leq N_R$)}
\end{equation}
where $N_R=120$ is the number of radial grid points. 

We set the polar angular grid to follow the scale height of the disc model assuming  irradiation \citet{Cunningham1976}, with $N_{\rm \theta} =240$ as the number of grid points defined on the angle from the mid-plain $\alpha_j =\pi/2 - \theta_\mathrm{N_\mathrm{\theta}-j}, (0\leq j \leq N_\mathrm{\theta})$  
\begin{equation}
\alpha_j  =
\begin{cases}
     \arctan  \left\{f_d  (R_\mathrm{j}/R_\mathrm{\mathrm{out}})^{2/7} \right\}   ,&( {\scriptsize \text{$0\leq j \leq N_R$}}) \\
    \arcsin\left\{ \frac{1.0-\sin(\alpha_\mathrm{N_R})}{N_\mathrm{\theta}-N_R}(j-N_R)+\sin(\alpha_\mathrm{N_R}) \right\} , &({\scriptsize \text{$N_R < j \leq  N_\mathrm{\theta}$}})
\end{cases}
\label{eq:polar grid}
\end{equation}
where $f_d= 1.5 \times 10^{-3} \left(\frac{L}{L_\mathrm{\mathrm{Edd}}}\right)^{1/7} \left(\frac{M_c}{M_\odot}\right)^{-1/7}\left(\frac{R_\mathrm{out}}{R_\mathrm{g}}\right)^{2/7}$
is the shape (H/R) of the irradiated disc from \cite{Cunningham1976}, as recast in \cite{Kimura2019}. 

We use outflow boundary conditions for both the inner and outer boundary in the R direction, where the material freely can leave the simulation grid but cannot enter. 
In $\theta$ direction, we use the axially symmetric boundary at the rotational axis of the accretion disc at $\theta = 0$ and 
reflecting boundary at $\theta_{N_\theta} $.
At each time step, we update the density at disc surface at $i=j $ as derived from the 
critical value of the pressure ionisation parameter.
The complex shape of the heating/cooling equilibrium curve shown in Fig.
\ref{fig:sed_thermal} has multiple instability regions.
The first one reached by the radiation going into the photosphere from above is at the end of the warm stable branch, at $\log\Xi_c\sim 0.9$ and $\log T=5.6$. 
This corresponds to a standard ionisation parameter of $\xi_c\sim 200$, so the material already has substantial soft X-ray opacity, shadowing the regions below, whereas the heating/cooling curves are calculated assuming unattenuated radiation.
Thus we use the first instability to determine the base density of the corona 
$\rho_{bc}=\mu m_p L/(\xi_c R^2)\exp(-\tau) $, where $\tau$ is the integrated optical depth through the wind along a line of sight from the centre to the disc surface, which means the boundary of the disc surface is time dependent (see also \citealt{Tomaru2019}).

We also assume there is further attenuation of the direct 
irradiation in an inner static corona region, with  $\tau_0=\exp(2(1-(\alpha/\alpha_{c})^2)) $, where $\alpha_c = H_c/R_{\rm ia}$ and $R_{\rm ia}$ is the radius of inner corona given by 
\begin{equation}
R_\mathrm{ia}/R_\mathrm{IC}= 0.021\Bigl[ \frac{T_\mathrm{IC,8} (L/L_\mathrm{Edd}) }{\Xi_{\rm H,min} 0.1} \Bigr]^{1/2}
\end{equation}
\citep{Begelman1983b} and $H_{c}= [2R^3/R_\mathrm{IC}]^{1/2}$ is its scale height. 
The $\Xi_{\rm H,min}= 190~(\log \Xi_{\rm H,min}=2.27)$ is the pressure ionisation parameter where $T=T_{\rm IC}/2$ on thermal equilibrium curve (right in Fig.\ref{fig:sed_thermal}) . 
This optical depth becomes smaller than unity when $R$ is larger than $0.06 R_{\rm IC}$.

As initial conditions in the disc region, $R_{i}\geq R_{j} $, we set velocity as $v_R =0$, $v_{\phi} = v_K$ and density as $\rho_{bc}$. In the non-disc region, 
we assume very small density 
$\rho =1.0 \times 10^{-30}~{\rm g~cm^{-3}}$, and 
$T_{0} =1.1 \times 10^7 (R/R_g)^{-3/4}{\rm K}$

\begin{figure}
    \centering
    \includegraphics[width=\hsize]{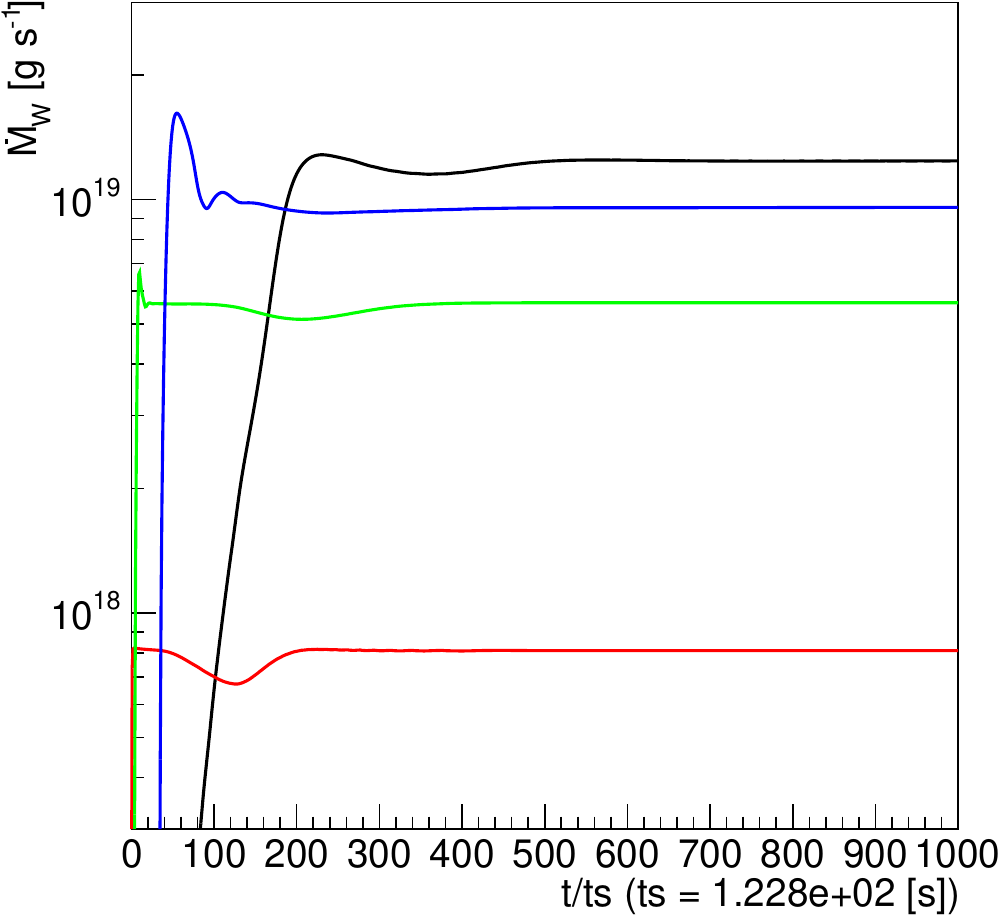}
    \caption{The mass loss rate at each spherical shells The colours shows $0.01 \leq R/R_{\rm IC}\leq 0.038$ (red), $0.038\leq R/R_{\rm IC}\leq 0.14$ (green),
    $0.14\leq R/R_{\rm IC}\leq 0.58$ (blue),
    and total mass loss rate at outer boundary (black). It is clear that the simulation has converged after the initial transients. }
    \label{fig:mass_loss}
\end{figure}

We run the simulation for about 3 sound crossing time $3R_{\rm IC}/c_{\rm IC}=1.2\times 10^{5}~{\rm s}$.
The mass loss rate of the simulation is converged after 1.5 sound crossing time (Fig.\ref{fig:mass_loss}). 
The mass loss rate is $1.24\times 10^{19}~{\rm g~s^{-1}} = \dot{M}_{\rm wind}/\dot{M}_{\rm a}=2.8$. 

The analytic thermal wind models predict $\dot{M}_{\rm wind}/\dot{M}_{\rm a} \propto \log (R_{\rm disc}/(0.1 R_{\rm IC}))$ for systems with the same $T_{\rm IC}$ and $L/L_{\rm Edd}$ (\citealt{Woods1996, Done2018}). 
This scaling works to compare the total mass loss rate to our previous 
simulation of the similar SED and luminosity source GX13+1 \citep{Tomaru2020b}, where $ \dot{M}_{\rm wind}/\dot{M}_{\rm a}\sim 8$ as the disc in GX13+1 has outer radius of $10R_{\rm IC}$ compared to $0.5R_{\rm IC}$ in \gro.

The thermal-radiative wind can in principle be launched from a smaller inner radius than $0.1~R_{\rm IC}$, because of the strong radiation force, but due to the shadowing of the inner corona (equation A3) out to $R\sim 0.1R_{IC}$, the amount of this inner wind is small. 
This smaller launching radius by radiation force is also shown in \citet{Proga2002}, though there are several 
differences in assumptions which mean that their simulations are not directly comparable to ours. 

\begin{figure*}
    \centering
    \includegraphics[width=\hsize]{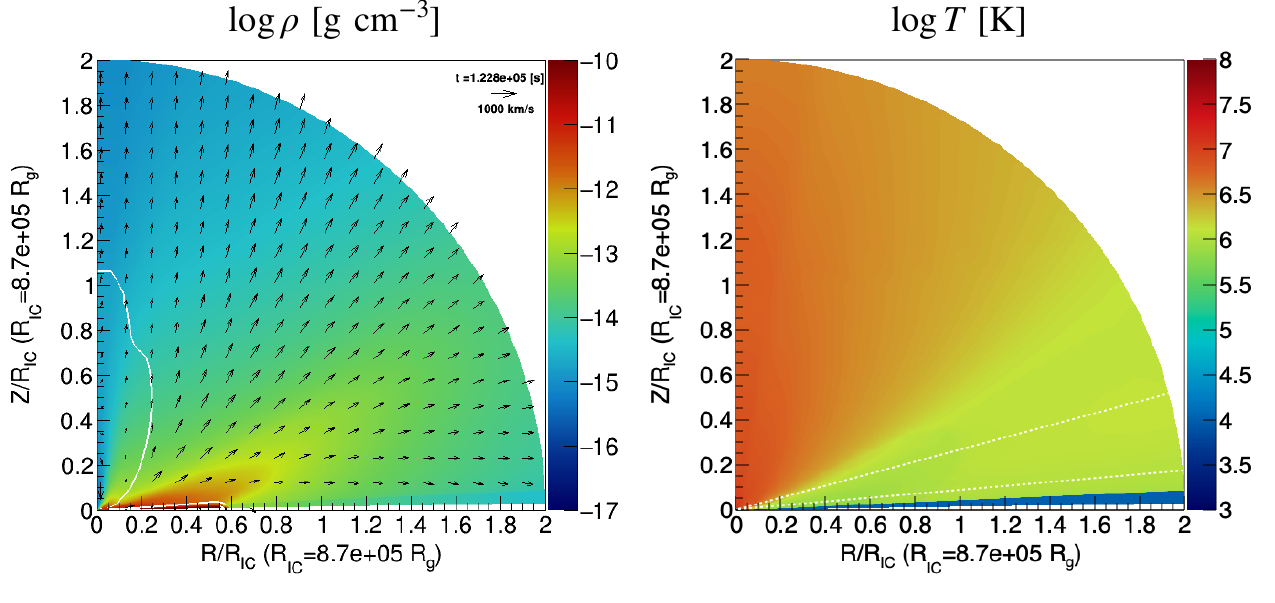}
    \caption{The distributions of density (left) temperature (right) obtained from our radiation hydrodynamic simulation. 
    The white line in left shows the Mach 1 surface.
    The white dashed line in the right shows the inclination angle of $75^\circ$ and $85^\circ$. 
    The region between these angle is   used for the plotting Fig. \ref{fig:R_on_thermal}. 
    }
    \label{fig:2d_plot}
\end{figure*}

\begin{figure*}
    \centering
    \includegraphics[width=\hsize]{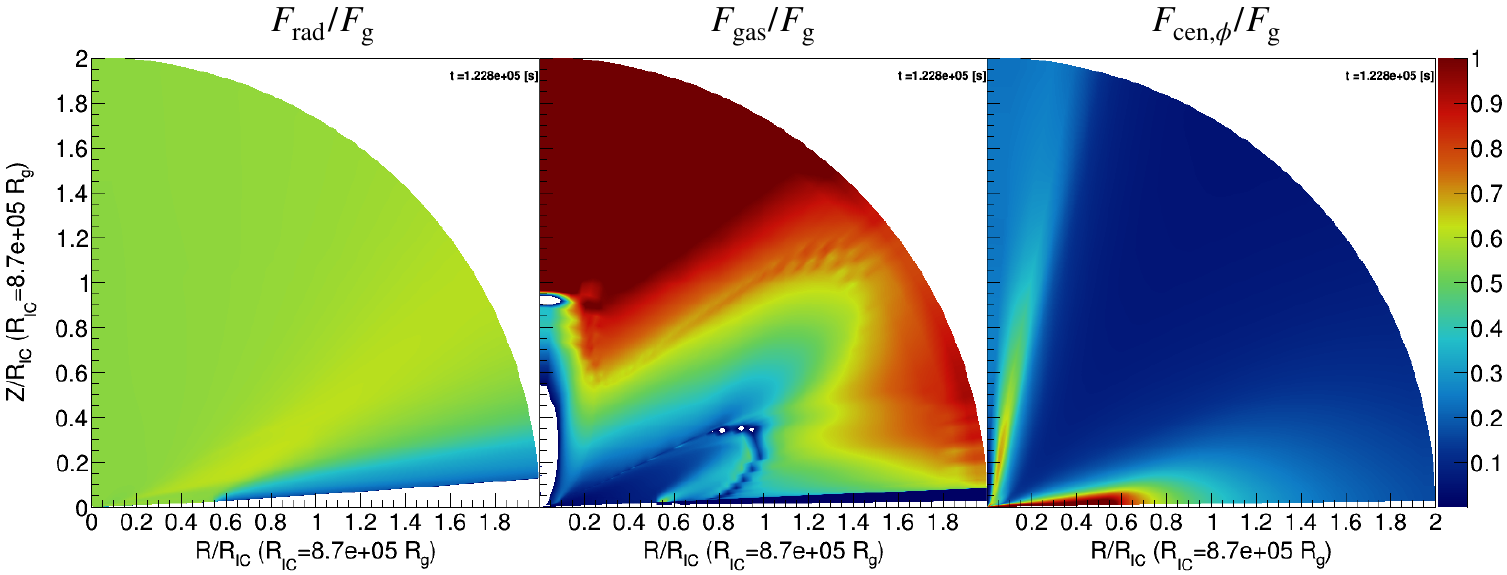}
    \caption{The distribution of radiation force (left),
    gas pressure force (middle),
    and centrifugal force (right), all given as a ratio to the local gravitational force.
    The radiation force is almost zero at high inclination angle with $R>R_{\rm disc}$ since the illumination is attenuated by disc structure.
    }
    \label{fig:force_2d}
\end{figure*}

We plot the final distribution of density and temperature (Fig.\ref{fig:2d_plot}).
These structures are similar to those in our previous simulation of GX13+1 \citep{Tomaru2020b} except for detailed parameters such as disc size (which also affects the simulation box size). The white blank region at mid-plane is out side of our simulation box. 
Fig. \ref{fig:force_2d} shows the distribution of radiation force (left), gas pressure force (middle), centrifugal force by $v_{\phi}$ (right), all given as a ratio to the local gravity. This shows that the
wind is driven by radiation force and centrifugal force at disc surface, but 
that gas pressure forces generally dominates at larger radii in the wind. 
At $R=0.3 R_{\rm IC}$, the radiation force is larger than 0.5, which means the bound-bound and bound-free are significant as well as free-free (electron scattering). 
This is consistent with a previous simulation \citep{Tomaru2019}
which also showed that electron scattering alone is not sufficient to get the observed velocity.

\begin{figure}
    \centering
    \includegraphics[width=\hsize]{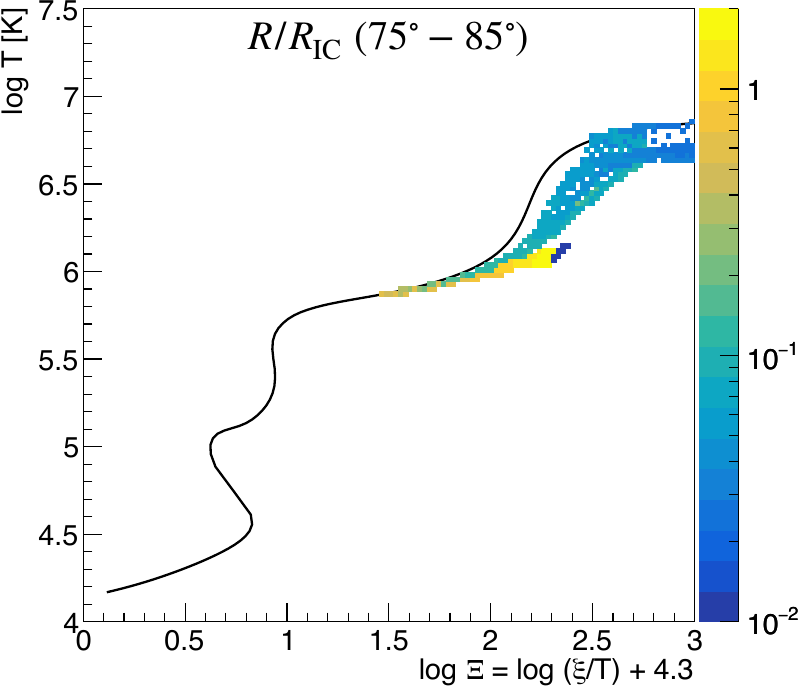}
    \caption{
    The black line shows the relation between pressure ionisation parameter and temperature for the thermal equilibrium curve for optically thin material. 
    The values for the wind material between $75^\circ-85^\circ$ are colour coded by their radius. 
    The higher ionised material is located at the inner radius ($\leq 0.2 R_{\rm IC}$, blue) , the lower ionised material is located at the outer radius ($> 0.2 R_{\rm IC}$, yellow). 
    Both are below the thermal equilibrium curve due to adiabatic losses.
    }
    \label{fig:R_on_thermal}
\end{figure}

We also plot the wind properties in $\Xi-T$ space between $75^\circ-85^\circ$ with radius (Fig.\ref{fig:R_on_thermal}).
The wind is converged to the thermal equilibrium curve at large radii, but 
the inner regions are slightly below this curve due to 
adiabatic cooling.
In the analysis, we calculate iron ion columns as $N_{\rm ion, Fe} = \sum \delta N_{\rm ion, Fe}=\sum f_{\rm ion, Fe}(\xi, T)A_{\rm Fe}\rho/(\mu m_p) \delta R$, where $f_{\rm ion, Fe}(\xi, T)$ is ion fraction, which is also calculated by {\sc cloudy}, and $A_{\rm Fe}=3.3\times 10^{-5} $ is the iron abundance with respect to hydrogen. 
We also calculate the average velocity weighted by those ion columns as $v_{\rm R, ion} = \sum \delta N_{\rm ion, Fe} v_{\rm R}/N_{\rm ion, Fe} $. 
These results are given in Fig.\ref{fig:integral} in the main paper.

We test that the numerical grid has sufficient resolution by using a smaller
$(N_r, N_\theta) = (90, 180)$, and larger $(N_r, N_\theta) = (240, 480)$ grids. 
In the smaller number case, the simulation is not well calculated at larger radius because the larger radius is sparse as grid is logarithmic. 
But the integral values (velocity and ion columns) are similar to the original simulation because these  values are mainly set by the inner region. 
The higher resolution grid model gives the same result as original model. 
Thus, we conclude our simulation grid is sufficient to capture the wind properties.

We checked also the impact of the 
inner boundary condition of the grid. The velocity structure of 
Fe\,{\sc xxvi} at high inclination changes
when the inner boundary condition is 
changed to  inflow/outflow instead of just outflow. 
However, the mass loss rate and ion columns are almost the same. 
So, the overall feature of the wind itself is not so different. 


\bsp	
\label{lastpage}
\end{document}